\documentclass[showpacs,amsmath,amssymb,prb]{revtex4}

\usepackage{graphicx,epsfig}
\usepackage{bm}
\usepackage{rotating,texdraw}

\begin{document}
%\preprint{APS/123-QED}

\title{Biphonons in the Klein-Gordon lattice}

\author{Laurent Proville}
\affiliation{\small
Service de Recherches de M\'etallurgie Physique, CEA/DEN/DMN Saclay
        91191-Gif-sur-Yvette Cedex, France}

\date{\today}

\begin{abstract}

A numerical approach is proposed to compute the phonon bound states
in a quantum nonlinear Klein-Gordon lattice. In agreement with
other studies\cite{AGRA,Eilbeck} on a different quantum lattice,
nonlinearity is found to lead to a phonon pairing and consequently
some biphonon excitations. The energy branch and the correlation
properties of the Klein-Gordon biphonon are studied in detail.
\end{abstract}


\maketitle

% *************************
\section{Introduction}
% *************************
In lattices made of identical particles,
the energy is formulated by the Hamiltonian operator:
\begin{equation}
H=\sum_l [\frac{p_l^2}{2m} + V(x_l) + \sum_{j=<l>} W(x_l-x_{j})].
\label{DebyeModif}
\end{equation}
where $x_l$ and $p_l$ are displacement and momentum of the
particle at site $l$, in a d-dimensional lattice. From the left to
the right hand side of the equation Eq.\ref{DebyeModif}, the
energy contributions are identified as the kinetic energy, the
local potential and the interaction between particles. Our purpose
is to study the case of quantum particles that are weakly
interacting, i.e., the onsite energy $V$ dominates the interaction
$W$. Physically, this can account for the coupling of internal
modes in molecular crystals. For small amplitudes of $x_l$, the
well-known harmonic approximation reduces $H$ to a sum of
quadratic terms, i.e., the linear Klein-Gordon (KG) Hamiltonian.
So, the Schr\"{o}dinger equation can be solved analytically. The
elementary excitation is a plane wave called an {\it optical
phonon} and whose energy is fixed by the wave momentum $q$, in the
lattice Brillouin zone. A consequence of the ideal harmonicity is
that higher order excitations are simply the linear superpositions
of these optical phonons.

For larger displacements, some non-quadratic contributions are
involved in the expansion of $H$. Then the nonlinear KG
Hamiltonian can no longer be diagonalized analytically. In a
nonlinear KG lattice, F. Bogani\cite{Bogani} derived some one- and
two- phonon renormalized Green functions and showed that the
nonlinear terms involve a pairing of the optical phonon modes. It
confirmed the existence of biphonon excitations, studied earlier
in a different lattice model by V.M. Agranovich\cite{AGRA1970}. A
convincing agreement was found between theory\cite{Bogani} and
experiments\cite{Bogani2,Califano} in molecular crystals where the internal
molecule bonds yield a strong nonlinearity. The direct
diagonalization of a KG Hamiltonian is, in principle, more precise
than the computation of Green functions since it requires fewer
approximations. In Ref.~\onlinecite{bishop96}, by treating
numerically the KG model, W. Z. Wang {\it and al.} confirmed the
existence of phonon bound states. Furthermore, some of these
states have been shown to feature a particle-like energy band, for
certain model parameters. The authors identified these specific
excitations as being some quantum breathers (see Refs.
\onlinecite{MacKay2000,Sievers2002,Flach2002,Fleurov,Eilbeck} for
more details about quantum breathers) because of their
counterparts in classical mechanics\cite{Siev,MA94}. Nonetheless,
the approach proposed in Ref.~\onlinecite{bishop96} requires a
huge computing cost so the size of lattices was limited to a
one-dimensional (1D) chain of 8 unit cells. Moreover, the
numerical simulations were restricted to the parameter region
where the non-harmonic part of the lattice energy is modelled by a
quartic onsite potential, i.e., the well-known $\phi^4$ model. In
the present paper, we propose a numerical treatment of the
nonlinear KG lattice which takes advantage of the weak coupling
($W$ in Eq.\ref{DebyeModif}). That permits to analyze lattices
large enough to approach the infinite system features and to study
different types of nonlinearity, as well as the two-dimensional
(2D) KG lattice.

We confirm that when nonlinearity is significant, a pairing of
optical phonon states occurs and the so-called
biphonon\cite{AGRA} branch contributes to the energy-spectrum.
That branch splits from the two-phonon band by opening a gap. The
width of that gap indicates the magnitude of nonlinearity since
the biphonon gap vanishes completely for a pure harmonic lattice.
In between the two types of lattice, i.e., harmonic and strongly
nonlinear, the binding energy of the biphonon drops to zero at the
center of the lattice Brillouin zone (BZ) while at the edge, the
biphonon excitations are still bound. Then, in the
energy-spectrum, the biphonon gap vanishes at the center, whereas
a pseudogap is found to open at the edge of BZ. We predict that
the pseudogap is a systematic feature of lattices in which
nonlinearity is moderate, whatever is the lattice dimension or the
dominant nonlinearity, i.e., $\phi^3$ or $\phi^4$. In addition, we
enhance how quantum properties of the biphonon depends on the
nonlinearity. When the biphonon gap opens, the Klein-Gordon
biphonon excitations show a finite correlation length, for all
momentum $q$, under the condition that the non-quadratic energy term
is a $\phi^4$ potential. That agrees with findings of
Ref.\onlinecite{bishop96}. Considering the cubic term in the
potential energy $V$, it involves a long range correlation of the
biphonon states. The space correlation properties of triphonon are
also studied and our results are used to establish some
expectations on the existence of breather-like excitations in the
quantum KG lattice.

The present paper is organized as follows. In Sec.\ref{Computing}
the model for the nonlinear discrete lattice is introduced and the
computing method is detailed. Then it is tested for the quadratic
lattices as well as for the $\phi^4$ model. In Sec.\ref{results}
our results are presented concerning the biphonon spectrum while
in  Sec.\ref{Correlations} the space correlation properties of the
phonon bound states are studied. Finally, these results are
discussed in Sec.\ref{Conclusion}.

\section{Model and computing method}
\label{Computing}

In Eq.\ref{DebyeModif}, at node $l$ of a translational invariant
d-dimensional lattice, the quantum particle of mass $m$ evolves in
a local potential $V$, being coupled to its nearest neighbours, $j$
by the interaction $W$. For moderate amplitudes of mass
displacements around equilibrium, $V$ and $W$ can be expanded as
Taylor series. The expansion of $V$ is truncated to the fourth
order $V(x_l)=a_2 x_l^2+a_3 x_l^3 +a_4 x_l^4$ while for $W$, only
the quadratic term is retained, $W(x_l-x_{j})=-c(x_l-x_{j})^2$.
Higher order terms can be treated with no difficulty in what
follows. Actually, they are found not to change qualitatively the
results, at least for reasonable values of energy coefficients,
consistent with optical modes.

Introducing the dimensionless operators $P_l = p_l / \sqrt{m
\hbar\Omega}$, $X_l=x_l \sqrt{m\Omega/\hbar}$ where the frequency
$\Omega = \sqrt{2 (a_2-2.c.d)/m}$ is defined for either a chain
$d=1$ or a square lattice $d=2$, the Hamiltonian reads
\begin{equation}
H = \hbar \Omega \sum_l \frac{P_l^2}{2} + \frac{X_l^2}{2} + A_3
X_l^3 + A_4 X_l^4 + \frac{C}{2}  X_l  \sum_{j=<l>} X_{j}
\label{Hamilton}
\end{equation}
where one finds the dimensionless coefficients:
\begin{eqnarray}
A_3= a_3 \sqrt{\frac{\hbar}{m^3 \Omega^5}}\text{,} \ A_4= a_4
\frac{\hbar}{m^2 \Omega^3} \ \text{and} \ C= \frac{4 c}{m
\Omega^2}\text{.} \nonumber
\end{eqnarray}

For the harmonic lattice, i.e., $A_3=0$ and $A_4=0$, the Fourier
transform of both the displacements, $X_l=\frac{1}{\sqrt{S}}\sum_q
e^{-i q\times l} \tilde{X}_q$ and the momenta,
$P_l=\frac{1}{\sqrt{S}}\sum_q e^{-i q\times l} \tilde{P}_q$
simplifies $H$ into a sum of independent Hamiltonian:
\begin{equation}
h_q=\hbar \Omega (\frac{\tilde{P}_q^2}{2} + \frac{{\omega_q}^2}{2} \tilde{X}_q^2)
\label{hq}
\end{equation}
with $\omega_{q}= \sqrt{1 - 2 C \sum_k cos{(q_k)}}$ and $q_k$ is
the dimensionless coordinate of the wave vector in the $k^{\text{th}}$
direction of the lattice. The periodic boundary conditions impose
$q_k=2\pi l_k/L_k$ where $L_k $ is the number of sites in the
$k^{\text{th}}$ direction and $l_k$ is an integer $l_k \in
[1,L_k]$. The lattice size is denoted $S=\Pi_{k=1}^{d} L_k$. Using
the standard harmonic oscillator theory, one finds the
eigenvalues:
\begin{equation}
\Lambda_{\{n_q\}}=\hbar \Omega \sum_q (n_q+\frac{1}{2}) \sqrt{1 +
2C\sum_k cos{(q_k)}}.\label{HarmoEigenV}
\end{equation}
The $n_q$ are quantum numbers that range from $0$ to infinity
and fix the energy contribution of the mode $q$. For the
nonlinear case, the aforementioned procedure is no longer simple
because non-quadratic terms yield a coupling between the $h_q$
operators. Indeed, writing the nonlinear energy term $(\sum_l
X_l^3)$ as a function of the displacements Fourier transform
$\tilde{X}_q$, gives $(\frac{1}{\sqrt{S}} \sum_{q,q'}
\tilde{X}_{q} \tilde{X}_{q'} \tilde{X}_{(-q-q')})$. The
computation of the corresponding bracket thus scales as $S^2$
as the sum runs over 2 wave vector,
while for the quartic term it would scale as $S^3$. Such a task has
been achieved in Ref.\onlinecite{bishop96} for the quartic term.
In the algorithm which follows, the computation of the brackets in
Eq.\ref{bra}, scales as $S$ which requires much less computation
time for a given $S$. Starting from the exact diagonalization of
the Hamiltonian where no interaction couples displacements, the
low energy states are used to construct a set of Bloch waves upon
which is expanded the entire Hamiltonian, including the coupling
$W$. The Schr\"{o}dinger equation is then approximately solved with
an error which shrinks to zero by increasing the basis cutoff.

The starting point is thus the eigenvalue problem for a single
oscillator $h_l=\hbar \Omega (\frac{P_l^2}{2} + \frac{X_l^2}{2} +
A_3 X_l^3 + A_4 X_l^4)$. The Bose-Einstein operators
$a_l^{+}=(X_l- iP_l)/\sqrt{2}$ and $a_l= (X_l + iP_l)/\sqrt{2}$ are
introduced in the writing of $h_l$:
\begin{eqnarray}
h_l=& &\hbar \Omega [a_l^+ a_l + \frac{1}{2} + \frac{A_3}{\sqrt{8}}
(a_l^{+3}+a_l^{3}+3a_l^{+}a_l^{2} +3a_l^{+2}a_l +3 a_l^{+}+3a_l)\nonumber\\
&+& \frac{A_4}{4} (a_l^{+4}+a_l^{4}+4a_l^{+3}a_l
+4a_l^{+}a_l^{3}+6(a_l^{+2}a_l^{2}+a_l^{+2}+a_l^{2}+2a_l^{+}a_l)+3)].
\label{DBa}
\end{eqnarray}
Expanding the operator $h_l$ on the Einstein states, i.e.,
$|n,l>=\frac{1}{\sqrt{n!}} a_l^{+ n} |\emptyset_l>$ for all $ n
\in \{0... N-1\}$ gives a matrix ${\cal M}$ of rank $N$. In each
row of ${\cal M}$, one finds the nonzero coefficients:
\begin{eqnarray}
{\cal M}_{n,n}&=&\frac{3}{2} A_4 n^2  + n (3 A_4+1)+\frac{3}{4} A_4 + \frac{1}{2} \nonumber\\
{\cal M}_{n,n+4}&=&\frac{1}{4} A_4\sqrt{(n+4)(n+3)(n+2)(n+1)} \nonumber\\
{\cal M}_{n,n+3}&=&\frac{1}{\sqrt{8}} A_3\sqrt{(n+3)(n+2)(n+1)}  \nonumber\\
{\cal M}_{n,n+1}&=&\frac{3}{\sqrt{8}} A_3 \sqrt{(n+1)^3}  \nonumber\\
{\cal M}_{n,n+2}&=& \frac{1}{4} A_4 (4n+6)\sqrt{(n+2)(n+1)} .
 \end{eqnarray}

When the cutoff $N$ tends to infinity, the Einstein states form a
basis in the space of onsite states. The Schr\"{o}dinger problem
for the Hamiltonian $h_l$ is thus equivalent to the
diagonalization of ${\cal M}$. That diagonalization is compared to
the semi-classical quantization\cite{Landau} in Fig.
\ref{fig1}(a), for the case of a He atom embedded into a
double-well potential: $V(x)=16 E_d/b^4 x^2 (x - b)^2$. The
parameters, $E_d$ and $b$, are the energy barrier and distance
between minima, respectively 
(see Ref. \onlinecite{Param1}). The
very good agreement proves the efficiency of the diagonalization
method even for a non-monotonic onsite potential. Arranging the
onsite eigenstates in increasing order of their eigenvalues, the
$\alpha^{\text th}$ eigenstate is denoted $\phi_{\alpha,i}$ and
its eigenvalue is $\gamma(\alpha)$. As shown in Fig.\ref{fig1}(b),
each eigenvalue $\gamma(\alpha)$ is found to converge to a steady
value as $N$ increases. In Fig.\ref{fig1} (b), the graphic does
not allow to distinguish $N=\infty$ from $N>100$. With today's
computers, the cutoff has been easily increased to $N=2000$ which
has been taken for the limit $N=\infty$ in Fig.\ref{fig1}(b). Increasing the
cutoff to values larger than $N=100$ does not change
significantly our final results on the low energy excitations of
the KG lattice (Sec.\ref{results}). In what follows, $N$ is
cautiously fixed to $N=500$ and then the time requirement to
diagonalize ${\cal M}$ is about few minutes on a PC computer. Note
that increasing $N$ implies no overload of the calculations
in the second part of the algorithm.

%In Fig. \ref{fig2}, for different sets of
%coefficients $A_3$ and $A_4$, the bracket of the displacement
%operator, $<\phi_{\alpha,i} |X_i|\phi_{0,i} >$, is plotted versus
%the rank $\alpha$ of $\phi_{\alpha,i}$. For further discussions
%(in Sec. \ref{Correlations}), it is noteworthy that the absolute
%value of this bracket is maximum at $\alpha=1$ and drops to zero
%with increasing $\alpha$. In addition,
%$<\phi_{\alpha,i}|X_i|\phi_{0,i} >$ takes zero value only for even
%$\alpha$'s, provided that $A_3=0$.\\
We now treat lattices with non-zero inter-site coupling. The
onsite state products $\Pi_i \phi_{\alpha_i,i}$ form a complete
orthogonal base for the lattice states. In order to reduce the
computer memory requirement, one takes advantage of the
translation invariance by introducing the Bloch wave formulation
for the state products. Among those states some equivalence classes
can be constructed in which each state results from a translation
applied to another state of the same class. Retaining only one
element for each translation class, the state which represents the
class is identified by the series of its $\alpha_i$'s, that is
denoted $[\Pi_{i} \alpha_{i}]$. The construction of the equivalence
classes is performed numerically. For each class, a Bloch wave can
be written as follows:
\begin{eqnarray}
B_{ [ \Pi_i \alpha_i]} (q) = \frac{1}{\sqrt{A_{ [ \Pi_i
\alpha_i]}}} \sum_j e^{-i q.j} \Pi_i
\phi_{\alpha_i,i+j}\label{OSPBW}
\end{eqnarray}
where $A_{ [ \Pi_i \alpha_i]}$ ensures the normalization. Some
attention must be paid to the possible translation symmetry of
the state products that may be higher than the lattice symmetry.
Indeed, for a given product there may be a lattice vector $t$ that
verifies $\Pi_i \phi_{\alpha_i,i}=\Pi_i \phi_{\alpha_i,i-t}$ with
coordinates $t_k$ such as $t_k<L_k$. It implies that $A_{ [ \Pi_i
\alpha_i]}=S.\Pi_{k=1}^d (L_k/t_k-fc(L_k/t_k))$ where
$fc(L_k/t_k)$ is the fractional portion of the ratio $L_k/t_k$.
Then the Bloch wave can only take the momentum $q$ such as
$q_k = 2 \pi p_k / L_k = 2 \pi p'_k / t_k$ where $p_k$ and $p'_k$
are some different integers.
%One could choose to
%work with $L_k$ as a prime number which
%reduces the high symmetry states to the uniform products $\Pi_i
%\phi_{\alpha,i}$ where $\alpha$ is constant.
The set of states $\{ B_{ [\Pi_i \alpha_i] }(q) \}_{q,N_{cut}}$,
including the uniform state $ \Pi_{i}\phi_{0,i} $ at $q=0$, form a
truncated basis where $N_{cut}$ fixes the upper boundary on the
onsite excitations: $\sum_i \alpha_i \leq N_{cut}$. When $C$ is
negligible, these states are the eigenstates of $H$. For
moderate values of $C$, they should be good approximates.
%As seen previously, for the single site problem, the truncation of
%the basis with a high enough cutoff may be expected not to affect
%the calculation of the low energy eigenstates. That assumption is
%verified further.
%for the harmonic lattice by comparing our calculations to
%the exact formula Eq. (\ref{HarmoEigenV}) (see Fig. \ref{fig3} and
%below in the text for more details).
%As the study is focussed on
%the low energy excitations of large size lattices, the uniform
%states $\psi_\alpha= \Pi_{i} \phi_{\alpha,i}$ with $\alpha>0$
%are not considered into the calculation since the difference
%$<\psi_\alpha|H|\psi_\alpha> - <\psi_0 |H| \psi_0> $ increases
%linearly with $S$ (provided the groundstate of $h_l$ is not degenerate).
Since the Bloch waves with different $q$, are not hybridized by
$H$, the Hamiltonian can be expanded separately for each
$q$. It is performed analytically and gives a matrix $\cal B$$(q)$
the coefficients of which are written as follows:
\begin{eqnarray}
<B_{ [ \Pi_i \alpha_i] }(q)| H |B_{ [\Pi_i \beta_i]}(q)>=& &
\frac{1}{\sqrt{A_{ [\Pi_i \alpha_i]} A_{ [\Pi_i \beta_i]}}} [
\Pi_i \delta_{\alpha_i,\beta_i} \sum_i \gamma(\alpha_i)
- \frac{C}{2} \sum_{l,j} \exp{(-i q \times j)}\nonumber\\
& & \sum_{k=<l>} D(\alpha_l,\beta_{l+j})
D(\alpha_{l+k},\beta_{l+k+j}) \Pi_{i\neq l,l+k}
\delta_{\alpha_i,\beta_{i+j}} ]\label{bra}
\end{eqnarray}
where $D(\alpha_l,\beta_l)$ denotes the bracket
$<\phi_{\alpha_l}|X_l|\phi_{\beta_l}>$ that is given by:
\begin{eqnarray}
D(\alpha_i,\beta_i)=\frac{1}{\sqrt{2}} \sum_{l=0}^N <\phi_{\alpha
,i}| l,i > (\sqrt{(l+1)} < l+1,i | \phi_{\beta ,i}> + \sqrt{(l)} <
l-1,i | \phi_{\beta ,i}>). \label{brk}
\end{eqnarray}
The eigenvalues of $\cal B$$(q)$ are computed numerically with an
exact HouseHolder method \cite{NR}.
%The only value of $N_{cut}$
%fixes the precision of our calculation
%which can be compared with
%the analytical expression in Eq. (\ref{HarmoEigenV}), for a pure
%harmonic lattice. For a one-dimensional (1D) chain, that
%comparison is done in Fig. \ref{fig3} where are reported our
%results for different $N_{cut}$. The insert in Fig. \ref{fig3}
%shows the two-phonon energy region.
%%%%%%%%%%%%%%%%%%%%%%%%%%%%%
In Fig.~\ref{fig4}, for a 1D chain $S=4$, our calculation is
compared to Ref.~\onlinecite{bishop96} (see
Ref.~\onlinecite{Param2} for conversion of model parameters). A
very good agreement is noted for the low energy states since the
eigen-spectra are superposed in Fig.~\ref{fig4}. In
Ref.~\onlinecite{bishop96}, the Schr\"{o}dinger equation was solved
by diagonalization of the matrix obtained from expanding $H$ in
the Einstein phonon basis, i.e., the eigenstates of the pure
harmonic lattice (see Eq.\ref{hq}). According to the authors :
``it restricts the numerical simulations to a parameter region
where nonlinearity is not too large''. In contrast, thanks to
the first step of our algorithm which solves the single site
nonlinear eigenvalue problem, we can treat all types of
nonlinearity (weak or strong and with $\phi^3$ or $\phi^4$ terms),
provided the inter-site coupling is not too large. For instance,
the approach of Ref.\onlinecite{bishop96} requires some
computations even for $C=0$ if $A_3\neq 0$ or $A_4\neq 0$, which
is straightforwardly solved in the first step of the present
algorithm.  Within the second step, in order to estimate the
accuracy of our calculations, different lattice sizes have been
tested for a reasonable value of $C$ ($C=0.05/d$, see Sec.
\ref{results}) and different $A_3$ and $A_4$. For small sizes,
$N_{cut}$ can be stepped up sufficiently to make eigenenergies
converge to steady values. The convergence is as fast as $C$ is
small, i.e., when $C=0$ the computation is exact (to machine
precision) and near instantaneous whereas when $C$ is raised, the
accuracy becomes worse because of inter-site coupling terms:
$(a_i^+ a_j^+)$ and $(a_i a_j)$ that involve hybridization with
high energy Bloch waves (Eq.\ref{OSPBW}), above the cutoff. Once
$N_{cut}$ has been determined to achieve the required precision,
then the lattice size is increased up to the capacity of our
computer memory. For instance, in a 1D lattice with $S=17$,
$N_{cut}$ has been varied from $N_{cut}= 3$ (68 Bloch waves,
Eq.\ref{OSPBW}) to $N_{cut}= 6$ (5940 Bloch waves). For $N_{cut}=
4$, the error on the low energy eigenvalues, say the 2 phonon
states, is inferior to $1 \%$ in comparison with $N_{cut}= 6$. The
size has thus been increased to $S=33$ (3052 Bloch waves) with no
noticeable discrepancy of the eigen-spectrum. For such a lattice,
the time required for the computation of the matrix $\cal B$$(q)$
scales in minutes whereas the diagonalization requires few hours
with a PC. This can be reduced to some minutes with a vectorial
computer and suitable numerical libraries. The matrices we have
to treat are much smaller than those in Ref.\onlinecite{bishop96},
which accounts for the tractability of our method. For $S=4$,
$65536$ states were required in Ref.\onlinecite{bishop96} whereas
only $19$ Bloch waves were required in our calculations for the
same lattice, with same parameters (see Fig.\ref{fig4}). This
increase in efficiency has been possible because we took advantage
of the weak inter-site coupling. For the 2D lattices, the size has
been limited to $S=13 \times 13$ which involves $4931$ Bloch waves
with $N_{cut}=3$. Some improvements are under
investigation. For example, the number of required states can be
reduced again by imposing $\alpha_i < n_{low}$ for Bloch waves
that verify $\sum_i \alpha_i> N_{low}$ in Eq.\ref{OSPBW}
($N_{low}<N_{cut}$). The integers $n_{low}$ and $N_{low}$ are then
adjusted so as not to change the precision over the low energy
eigenstates.

%This procedure is stopped either when
%the computer memories are overwhelmed or when a deviation larger than
%the required precision is observed in the eigen-spectra.
%These performances
%could be improved again by computing the $\cal B$$(q)$
%diagonalization with a non-exact numerical procedure which have
%not been done here.

%The use of Bloch waves that are introduced in the present section
%takes advantage of the translational symmetry and
%the matrix rank of eigenvalue problem is thus divided by a
%factor $S$, i.e., the lattice size.

%%%%%%%%%%%%%%%%%%%%%%%%%%%%%%%%%%%%%%%%%%%%%%%%%%%%%%%%%%%%%%%%%%%%%%%%%%%%%%%%%%%%%%%%%%%%%%%%%%%%
\section{Gap and pseudogap in the optical phonon spectrum}
%%%%%%%%%%%%%%%%%%%%%%%%%%%%%%%%%%%%%%%%%%%%%%%%%%%%%%%%%%%%%%%%%%%%%%%%%%%%%%%%%%%%%%%%%%%%%%%%%%%%
\label{results}

% Intro = first 1D, 2D at the end
For different values of nonlinear coefficients $A_3$ and $A_4$, we
first examine  the vibration spectrum of a one-dimensional (1D)
lattice. The 2D lattice is treated at the end of the present
section.
% what is the harmonic spectrum
When the non-quadratic part of the lattice energy is negligible,
the eigen-spectrum of $H$ is composed of the fundamental optical
branch due to the harmonic phonon states (in Eq.
\ref{HarmoEigenV}, a single $q$ verifies $n_q=1$) and the branches
due to the linear superpositions of these phonons (in Eq.
\ref{HarmoEigenV}, several $q$'s verify $n_q=1$). The latest
branches are stacked together into distinct bundles, each of them
filling in a compact range of energy. In Fig. \ref{fig5} and
following ones, each eigenvalue of the finite size Hamiltonian is
plotted as a single circle symbol. The distinct eigen-energies
participate in different branches. The phonon branch is marked
with the tag $\{1 \}$ while the branches that are due to the
linear superposition of 2 phonon states are labelled by the tag
$\{11\}$. For a macroscopic system, the bundle $\{11\}$ covers a
dense range of energy and forms a continuous band. The width of an
optical phonon branch being physically a few percent of the
elementary excitation energy, the dimensionless coupling $C$ is fixed
to $C=0.05$ which gives, indeed a phonon branch width $\Delta_1
\approx 10 \%$ of the phonon energy (see left inserts in Figs.
\ref{fig5}-\ref{fig6}).
% the gap
When nonlinearity is significant, the phonon branch shows no
qualitative change (left inserts in Figs.~\ref{fig5}(a) and
\ref{fig6}(a)) in comparison with the fundamental optical branch
in harmonic lattice. On the other hand, an isolated spectral branch
is found in addition to the phonon branch and its combination tones
(see Figs.\ref{fig5}(a) and \ref{fig6}(a) and right-hand inserts).
In Fig.\ref{fig7}, varying artificially the coupling $C$ from the
trivial case $C=0$, demonstrates that the additional branch
coincides with the energy of the Bloch wave $B_{[\alpha,0,...,0]}$
with a single onsite excitation $\alpha = 2$. In Fig.\ref{fig5}
and the next ones, the additional branch is marked with a single tag
$\{ 2 \}$. By analogy with biphonon theory\cite{AGRA}, this branch
is identified as the biphonon energy. Similar results are found
for the triphonon states whose branch is labelled by the tag $\{ 3
\}$ in Fig.\ref{fig7}. The reason for these isolated branches is
that onsite Hamiltonian eigenvalues $\gamma_{\alpha
>1}$ do not match the linear fit given by $(\gamma_1-\gamma_0)
\alpha+\gamma_0$. This is the consequence of $h_l$ anharmonicity.
The differences $\gamma_{\alpha>1} - [(\gamma_1-\gamma_0)
\alpha+\gamma_0]$  involve some gaps in the $C=0$ spectrum which
is composed of the Bloch wave energies. A moderate inter-site
coupling hybridizes these states Eq.\ref{OSPBW} but the largest gaps
remain (Fig. \ref{fig7}). The raising of degeneracy of $B_{[\Pi_i
\alpha_i]}$ where only $2$ $\alpha_i$'s equal $1$ and the rest are
zero (number of these states is $S(S-1)/2$, their energy is
$2\gamma(1)+(S-2)\gamma(0)$ at $C=0$) yields a bundle of branches
which correspond to the linear superposition of 2 single phonon
states. In Fig.\ref{fig7}, for higher energy, other bundles are
labelled out by the tags $\{111\}$ and $\{21\}$. At zero coupling,
these branches coincide with the energies of states
$B_{[1,1,1,0,...,0]}$ and $B_{[2,1,0,...,0]}$. For a macroscopic
lattice, the bundles $\{111\}$ and $\{21\}$ form some dense bands,
as well as $\{11\}$. They are the unbound associations of 3 phonon
states and of a biphonon with a single phonon, respectively.
% The gap
In Figs.\ref{fig5}(a) and \ref{fig6}(a), for different parameters,
the biphonon branch splits from the 2 phonons band. Measuring the
energy of a biphonon state with reference to the unbound 2 phonons
for same momentum $q$, a binding energy of biphonon is defined. A
positive binding energy occurs when the onsite potential $V$ is
harder than a harmonic function (Fig.\ref{fig5}(a)) whereas a
softening yields a negative binding energy (Fig.\ref{fig6}(a)).
The biphonon energy gap is determined as the minimum of the
absolute value of binding energy with respect to $q$. The biphonon
gap reveals the strength of nonlinearity since when the biphonon
gap overpasses the phonon branch width (as it does in Figs.
\ref{fig5}(a) and \ref{fig6}(a)), it clearly indicates a
significant contribution of non-quadratic terms.
% The pseudogap
While in Figs.\ref{fig5}(a) and \ref{fig6}(a), a biphonon gap
opens, it is found that when nonlinearity is weak the biphonon
binding energy vanishes at center of the lattice Brillouin zone
(BZ) (Figs.\ref{fig5}(b) and \ref{fig6}(b)). However, at the edge
of BZ, the binding energy is comparable to the width of the phonon
branch $\Delta_1$ (inserts in Figs.\ref{fig5}(b) and
\ref{fig6}(b)). Consequently, a pseudogap is yielded when
the non-quadratic energy has same magnitude as inter-site coupling. In
this regime, the biphonon excitations exist only at the edge of BZ
while they are dissociated into unbound phonons at center. With
similar results, other calculations have been performed for
different parameters. They showed that the biphonon pseudogap is a
generic feature of lattices where nonlinearity is moderate.
Similar pseudogaps have been noted
in different quantum lattices\cite{Papanicolaou,Pouthier,Dorignac}.
The pseudogap opens at the edge of BZ, even though the
coupling sign is changed. So it is the $q$-range where nonlinear
behavior is likely to be experimentally measured in materials
where nonlinearity is weak.

% TRiphonon
In Fig.\ref{fig7}, the low energy eigenvalues of $H$ are plotted
versus parameter $C$. The variations of the energy branches of the
nonlinear excitations are labelled by tags defined previously.
It can be noted that the widths of bound states branches increase
with $C$ much slower than unbound phonon bands. The branch of the
$\alpha$ phonon bound states (biphonon for $\alpha=2$ and
triphonon for $\alpha=3$) are found to merge with unbound phonon
bands for a certain threshold $C_\alpha$.  At $C<C_\alpha$, the
$\alpha^{\text th}$ branch and the unbound phonon bands are
separated by a gap whereas around $C\approx C_\alpha$, only a
pseudogap separates them partially. The $C_\alpha$ threshold
depends on both coefficients $A_3$ and $A_4$ and it is different
for each $\alpha$ phonon bound state branch because of
anharmonicity. A unique set of nonlinear parameters $A_3$ and
$A_4$ corresponds to the energy distribution of the biphonon and
triphonon branches. So in principle, if a spectroscopy is able to
measure the biphonon and the triphonon resonances, it is
sufficient for inverting our numerical treatment and thus
determining the nonlinear parameters. Moreover, the results in
Fig.\ref{fig7} demonstrate that even though the non-quadratic
terms in $V$ are not large enough to open a biphonon gap, i.e.,
$C>C_2$ then a gap or at least a pseudogap opens for the $\alpha$
phonon bound states with $\alpha>2$. Finally, the theoretical
results in Fig.\ref{fig7}(b) are qualitatively similar to the
experimental findings in Ref.\onlinecite{Mao} in which the Raman
analysis of a molecular $H_2$ crystal shows a pressure-induced
bound-unbound transition of the so called bivibron around $25$
GPa. There is indeed a likeness between Fig.$10$ in
Ref.~\onlinecite{Mao} and the 2 phonon energy region in
Fig.~\ref{fig7}(b). In our model, the pressure variation of
experiments\cite{Mao} can be simulated by a change of the coupling
parameter $C$ due to the fact that neighbouring molecules are
moved closer together because of the external pressure. Actually,
the increase of $C$ induces a bound-unbound transition of the
biphonon at $C=C_2$.\\

% 2D case
In Fig.~\ref{fig10}, the diagonalization of a 2-dimensional (2D)
lattice Hamiltonian is performed for $A_3=0$ and for different
values of $A_4$. The coupling amplitude is such as the phonon band
width $\Delta_1$ is a few percent of the elementary excitation
branch. By estimating that $\Delta_1 \approx 2.d.|C|$, the value
of the dimensionless coupling is fixed around $C=0.025$. In the
first overtone region, a gap opens when $A_4$ is large (top of
Fig.\ref{fig10}) whereas that gap closes at the center of the BZ
when $A_4=0.025$ (bottom of Fig. \ref{fig10}). In the latter case,
a pseudogap is found to open around $q= [11]$ and the width of
that pseudogap has same order of magnitude as the phonon band
width. These both quantities can be compared in the inserts of
Fig.\ref{fig10} where the spectrum profile along $[11]$ is
plotted. The pseudogap width is same as in Fig.\ref{fig5} for the
1D chain. Consequently, a pseudogap is expected for all lattice
dimensions when both the inter-site coupling and the non-quadratic
energy have a comparable magnitude.

\section{Correlation properties of the phonon bound states}
%%%%%%%%%%%%%%%%%%%%%%%%%%%%%%%%%%%%%%%%%%%%%%%%%%%%%%%%%%%%%%%%%%%%%%%%%%%%%%%%%%%%%%%%%%%%%%%%%%%%
\label{Correlations}

In the present section, the study is focused on the space
correlation in a 1D lattice. The space correlation function for
displacements is defined as follows:
\begin{equation}
f(\Phi,n)= \sum_l  <\Phi | X_l X_{l+n} |\Phi>-
<\Phi | X_l|\Phi> <\Phi | X_{l+n} |\Phi>
\end{equation}
where $\Phi$ is an eigenstate and $n$ is the dimensionless
distance ($n>0$). For an harmonic lattice, at weak inter-site coupling, 
it is easily verified 
that the single phonon correlation function is well
approximated by ($cos(q\ n)$). 
Then, the correlation functions exhibit a
space modulation with a constant amplitude. At $q=0$, the function
$f(\Phi ,n)$ is a non-zero constant for all $n$.

Before discussing our results, it is convenient to compute
analytically the function $f(\Phi ,n)$ for the Bloch waves
$B_{[\Pi_i \alpha_i]}$ (Eq. \ref{OSPBW}) with only one onsite
excitation of order $\alpha$. These states have the form:
\begin{eqnarray}
B_{[\alpha,0,...,0]} (q) = \frac{1}{\sqrt{S}} \sum_j e^{-i q.j}
\phi_{\alpha,k+j}. \Pi_{l\neq k} \phi_{0,l+j}.\label{OSPBW2}
\end{eqnarray}
Let us introduce the notation, $\psi_\alpha(q)$ for these states.
At the uncoupled limit ($C=0$), the onsite excitation Bloch waves
correspond to some eigenstates of $H$ (see Fig.\ref{fig7} and text
in Sec.\ref{results}). The space correlation function of
$\psi_\alpha(q)$ is given by:
\begin{equation}
f(\psi_\alpha(q) , n)= 2 D(\alpha,0)^2 \cos{(q\ n)} -
\frac{(D(\alpha,\alpha)-D(0,0))^2}{S} \label{EQW}
\end{equation}
where $D(\alpha,0)$ and $D(\alpha,\alpha)$  are defined in
Eq.\ref{brk}. The constant term in the right hand side of Eq.
\ref{EQW} shrinks to zero when the lattice size $S$ increases. In
the numerical computations that are presented below, for finite
lattices, that constant is removed in order to approach the
behavior of the infinite lattices. Another point which is
noteworthy is that, in general, the displacement operators are
correlated for a given state $\psi_\alpha(q)$. Indeed the bracket
$D(\alpha,0)$ is usually not zero, except for even values of
$\alpha$ when $A_3=0$, and the function $f(\psi_\alpha (q) ,n)$
shows a space modulation of amplitude $R_\alpha=2 D(\alpha,0)^2$.
The coefficients $R_\alpha$ are given in table Tab.\ref{tab} for
parameters of Figs.\ref{fig5} and \ref{fig6}.

% ************************************************************************************
% To show that R_\alpha fixes the long range modulations of the multi-phonon bound states
% ***********************************************************************************

The function $f(\Phi,n)$ is plotted in
Figs.\ref{fig12}-\ref{fig13}  for eigenstates $\Phi$ that are
either the phonon, biphonon or the triphonon states. For each of
these states, $f(\Phi,n)$ is computed for several values of $q$.
The calculations at $C>0$, demonstrate that the long range
behavior of the correlation function for the $\alpha$ phonon bound
states ($1<\alpha<4$) as well as for the phonon states
($\alpha=1$) is similar to the variations of the function
$f(\psi_\alpha(q), n)$, i.e., the modulations have same amplitudes
for large distances. For a weak coupling, 
this property holds provided that the
$\alpha^{\text th}$ discrete branch is separated from the rest of
energy spectrum by some gaps.
% The phonon
In Figs.\ref{fig12}(a)-\ref{fig13}(a), the correlation function of
phonon states shows a modulation which extends over the whole
lattice as in the pure harmonic lattice. For large $n$, the
variations of $f(\Phi ,n)$ are same as ones of $f(\psi_1 (q), n)$,
Eq. \ref{EQW}. Indeed, the amplitude of the modulations is given
by the coefficient $R_1$ (see Tab. \ref{tab}). The correlated
character of phonon states is thus related to the feature of the
onsite excitation Bloch wave $\psi_1(q)$ and more precisely to the
fact that the onsite displacement operator generates an overlap
between the onsite groundstate $\phi_{0,i}$ and the first excited
state $\phi_{1,i}$. The agreement between the long range behavior
of phonon states and $\psi_1(q)$ holds in the harmonic lattice provided that 
$C$ is not too large.
This can be checked by comparing the formula $f(\Phi,n)=
cos(q\ n)$ for a single phonon state in the harmonic chain with
the equation Eq.\ref{EQW}, noting that for a quadratic form of
$V$, we have $D(1,0)=1/\sqrt{2}$ and thus $R_1=1$.

% The biphonon
The agreement between the coefficient $R_\alpha$ and the long
range modulations of $f(\Phi,n)$ is not strictly proved for
biphonon and triphonon (see Tab.\ref{tab} and
Figs.\ref{fig12}(b-c) and \ref{fig15}(b-c)) because the
corresponding branches are too close from the unbound phonon
bands. Nevertheless we estimate that the agreement roughly holds
%It can be noted in Figs.\ref{fig12}(b) and \ref{fig15}(b) that it
%does indeed
, whereas in Figs.\ref{fig13}(b)
%and \ref{fig14} (b)
it does not because the biphonon gap closes for small $q$.
Provided the $\alpha^{th}$ discrete branch is isolated in the
energy spectrum, there is no conceptual difference between
$\alpha$ phonon bound states and single phonon states because they
all come from the raising of the translational degeneracy of the
onsite eigenstates $\phi_{\alpha,i}$. Actually, we conclude that
the coefficient $R_\alpha$ is a good indicator of long range
correlations for the $\alpha$ phonon bound states (including
phonon states with $\alpha=1$) on the condition that the
hybridization with other phonon states is negligible. Therefore,
when $R_\alpha=0$, the $\alpha$ phonon bound states exhibit a
finite correlation length provided the $\alpha^{th}$ discrete
branch is isolated in the energy spectrum. In Fig.\ref{fig12}(b),
the space correlation function drops to zero for the large $n$ and
for all $q$ in agreement with $R_2$ equals zero (see
Tab.\ref{tab}).

As shown in Tab.\ref{tab}, $R_\alpha=2D(\alpha,0)^2$ decreases to
zero with increasing $\alpha$. Above $\alpha=4$, $R_\alpha$ may be
consider as physically negligible. By extension of our results
about space correlations of the $\alpha$ phonon bound states with
$\alpha<4$, it is expected that the $\alpha$ phonon bound states
with $\alpha \geq 4$ have a finite correlation length provided
their energy branch remains isolated from the linear combination
bands of the lower energy states, i.e., from the unbound phonon
states as well as from the combinations of the lower order
multi-phonon bound states (for instance, see $\{21\}$ in
Fig.\ref{fig7}). For the specific case $A_3=0$ and $A_4>0$, the
even values of $\alpha$ verify $R_\alpha=0$, so the corresponding
$\alpha$ phonon bound states have also a finite correlation length
when their branch is isolated in the spectrum.
% *******************************************************************
% Pseudogap = what happens at the gap closure:
% *******************************************************************
For a closing biphonon gap, a pseudogap was found to persist at
the edge of the lattice Brillouin zone (Sec.\ref{results}). In
Fig.\ref{fig13}(b), it is shown that the biphonon states have a
finite length scale in $q$-range where the
pseudogap opens. Indeed, due to hybridization with the 2 phonon
unbound states, the biphonons at $q \approx 0$ have a non-vanishing
space correlation function which goes beyond the finite
lattice size. \\

% *******************************************************************
% BREATHER
% *******************************************************************
The phonon bound states with a finite correlation length
exhibit a singular feature that is worth characterizing more precisely.
%For that purpose we focus on the biphonon since it
%can be computed with great accuracy. The results will be extended to
%the $\alpha$-phonon bound states.
For that purpose, denote by $\Phi_\alpha (q)$ the wave function
of the $\alpha$ phonon bound state with momentum $q$ and
by $E_\alpha(q)$ the dimensionless energy of the $\alpha^{\text th}$
branch. One introduces the time dependent Wannier state
$W_\alpha(t,n)$, which is constructed from a combination of the
$\alpha$ phonon bound states:
\begin{equation}
|W_\alpha(t,k)>= \frac{1}{\sqrt{S}}\sum_q e^{-i  (q \times k +
E_\alpha(q) \Omega t)} |\Phi_\alpha (q)>. \label{WTa}
\end{equation}
The subscript $k$ indicates the lattice site where is centered the
Wannier transform. In Fig.\ref{fig18}, the scalar
product $|<\psi_\alpha (q)|\Phi_\alpha (q)>|^2$ is plotted for
$\alpha=\{1,2,3\}$. Fixing $A_3=0$ and $C=0.05$, the parameter
$A_4$ is increased from zero. The scalar product $|<\psi_\alpha
(q)|\Phi_\alpha (q)>|$ is proved to equal $1$ when the gaps that
surround the $\alpha^{\text th}$ branch are large, or equivalently
when $A_4$ is large. Then, one can reasonably argue that the Bloch
wave $\psi_\alpha(q)$ is a very good approximate of the $\alpha$
phonon bound state and the corresponding eigen-energy is
approached by a perturbative theory:
\begin{equation}
<\psi_\alpha(q) |H|\psi_\alpha(q)>= \gamma_\alpha + 
(S-1)(\gamma_0 -C D(0,0)^2)-
C (R_\alpha cos(q) + D(\alpha,\alpha)D(0,0)).
\end{equation}
When $A_3=0$, the correlation function $f(\psi_2,n)$ is zero for
all $n$ (since $R_2=0$, see Tab.\ref{tab}), so for the biphonon,
we obtain $E_2 (q) =\gamma_2 + (S-1)\gamma_0 $. Then the energy
$E_2(q)$ does not depend on $q$, $E_2(q)=E_2$, and the Wannier
state $|W_2(t,k)>$ can thus be rewritten as follows:
\begin{equation}
|W_2(t,k)>=  e^{-i  ( E_2 \Omega t)}  \phi_{2,k} \Pi_{l\neq k} \phi_{0,i}. \label{WTa2}
\end{equation}
It is obvious that such a state is a localized and time periodic
excitation. The classical counterparts of such state would be the
breather solutions for the classical nonlinear discrete lattice,
e.g. A. J. Sievers and S. Takeno\cite{Siev}. These
classical breather solutions have two important features that are
first their spatial localization and second their time periodicity
with a frequency that is out of both the linear classical phonon
branch and its resultant harmonic bands (for an exact proof see
Ref.\onlinecite{MA94}). Our proposition about the classical
counterparts could be justified by comparing
the energies of a localized time periodic Wannier state and the
semi-classical quantization of the classical breather orbits, in
same lattice. In the simple case of zero inter-site coupling $C=0$,
such a comparison has been performed 
for the onsite double-well
potential (see Fig.\ref{fig1}(a) and text in Sec.\ref{Computing}).
The remarkable agreement obtained at $C=0$, allows to expect that 
our proposition 
holds for weak values of $C$, at least.
Extending our sketch to all $\alpha$ phonon bound states, the
conditions for a breather-like Wannier state in a KG lattice are
twofold: first, the branch of the $\alpha$ phonon bound states
must be isolated by large surrounding gaps and second,
$R_\alpha=0$. Let us detail the breaking of these conditions. When
the $\alpha^{\text th}$ branch is isolated but $R_\alpha > 0$, the
$\psi_\alpha (q)$ are still some good approximates for the
$\alpha$ phonon bound states (see in Fig.\ref{fig18} for the
phonon and triphonon). Then the band width of the $\alpha^{\text
th}$ branch does not vanish
%(see Fig. \ref{fig11})
because of the self-tunneling of $\psi_\alpha$, i.e., the
intersite coupling hybridizes the Bloch wave $\psi_\alpha (q)$
with itself. The Wannier transform (Eq.\ref{WTa}) of $\psi_\alpha$
is a combination of states non-coherent in time, since $E_\alpha
(q)$ depends on $q$. It is no longer a time-periodic solution. 
After a certain time $\Delta t_c$, the Bloch
waves $\psi_\alpha$ do no longer interfere. This $\Delta t_c$
varies inversely to the band width of the $\alpha^{\text th}$
branch. On the other hand, when $R_\alpha = 0$ but the
$\alpha^{\text th}$ branch is not isolated, the Bloch wave
$\psi_\alpha$ is hybridized with unbound phonon states. Then the
Wannier state in Eq.\ref{WTa} is no longer strongly localized as
the breather-like Wannier state in Eq.\ref{WTa2}, but it shows
some spacial extensions. In addition, $E_\alpha (q)$ depends on
$q$ (as shown in Fig.\ref{fig5}(b), for the biphonon). Hence the
Wannier transform is not coherent in time and $W_\alpha(t,k)$ is
not time-periodic.

%%%%%%%%%%%%%%%%%%%%%%%%%%%%%%%%%%%%%%%%%%%%%%%%%%%%%%%%%%%%%%%%%%%%%%%%%%%%%%%%%%%%%%%%%%%%%%%%%%
\section{Conclusion}
%%%%%%%%%%%%%%%%%%%%%%%%%%%%%%%%%%%%%%%%%%%%%%%%%%%%%%%%%%%%%%%%%%%%%%%%%%%%%%%%%%%%%%%%%%%%%%%%%%
\label{Conclusion}

In the present paper, the pairing of phonon states in the
nonlinear KG lattice has been considered with
numerics. First, the biphonon spectral features have been studied
and a pseudogap has been shown to characterize
lattices where nonlinearity is comparable to phonon band width,
whatever are the lattice
dimension and the type of nonlinearity.
Second, we have shown how
properties of the KG biphonon depend on the nonlinearity.
In the specific case of a strong $\phi^4$ nonlinearity, the biphonon
states feature a finite correlation length. The corresponding
Wannier transforms are breather-like excitations
that remain coherent for
a duration which increases with the strength of the $\phi^4$ nonlinearity.
In contrast, when a
$\phi^3$ nonlinearity predominates,
the biphonon exhibits a long range correlation
and biphonon tunneling is no longer negligible,
although it remains weaker than for single phonons. 
According to our results on the triphonon, one may expect that the
tunneling of the high energy multi-phonon bound states vanishes for any
type of nonlinearity, provided that
their energy branches avoid the unbound phonon bands. Then
some high energy breather-like excitations could occur.
These states should correspond to the semiclassical quantization
of the classical breather orbits.

I gratefully acknowledge financial support from Trinity
College and an EC network grant
on ``Statistical Physics and Dynamics of Extended Systems''
for the period spent at the University of
Cambridge when these ideas were developed. Many thanks are addressed
to Robert S. MacKay and Serge Aubry.

%Some experimental investigations on the biphonon pseudogap may
%help to quantify the nonlinear behavior of some common materials.

\newpage

\centerline{Table caption}

{Tab. \ref{tab}: Values of $R_\alpha=
2<\phi_{\alpha,i}|X_i|\phi_{0,i}>^2$ for different $\alpha$ and
different nonlinear parameters, $A_3$ and $A_4$ used in
Figs.\ref{fig5}-\ref{fig13}.}

\newpage

\centerline{Figure captions}

\begin{figure}
\noindent
\includegraphics[width= 12cm]{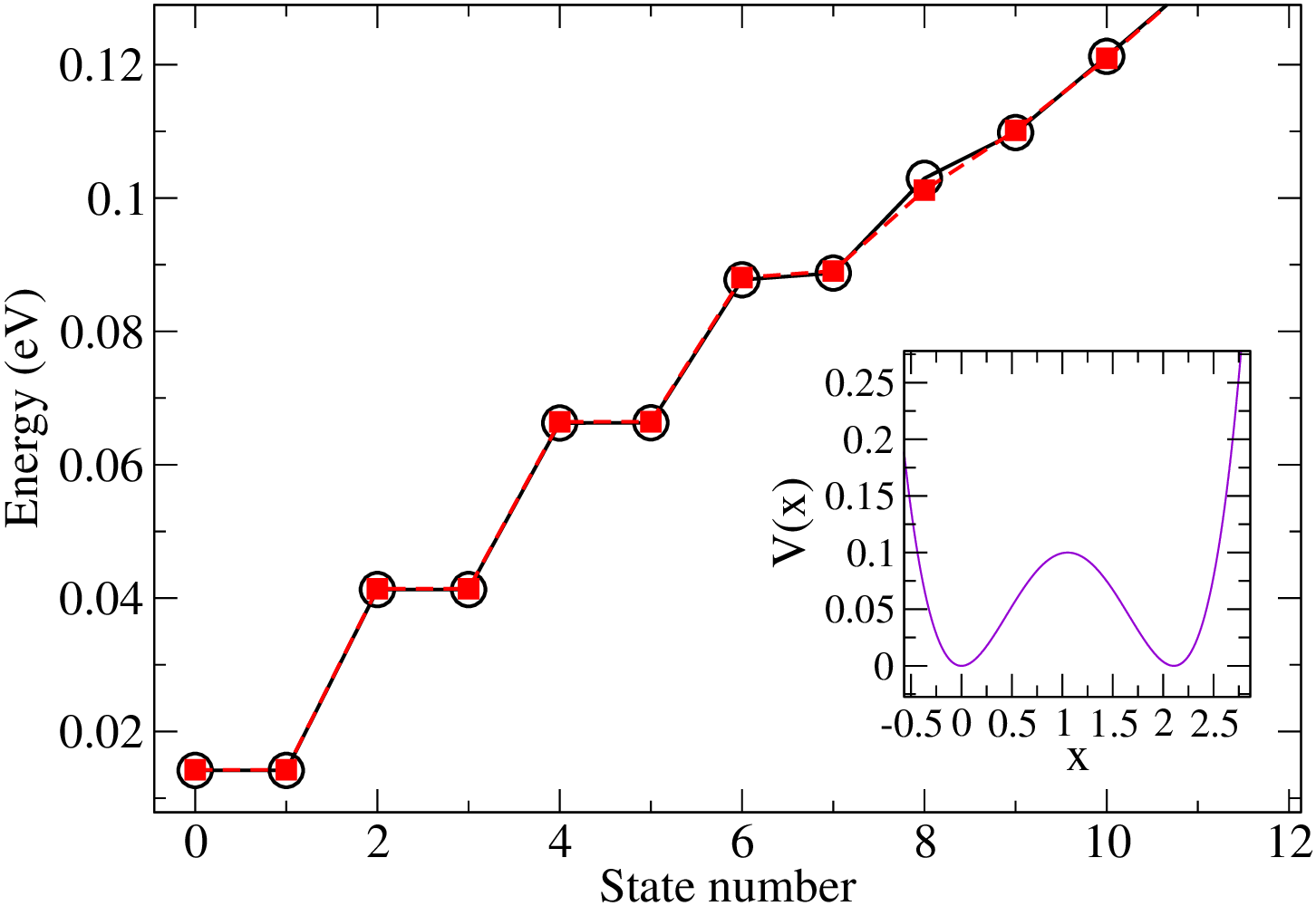}\\
\vspace{2.5cm}
\includegraphics[width= 12cm]{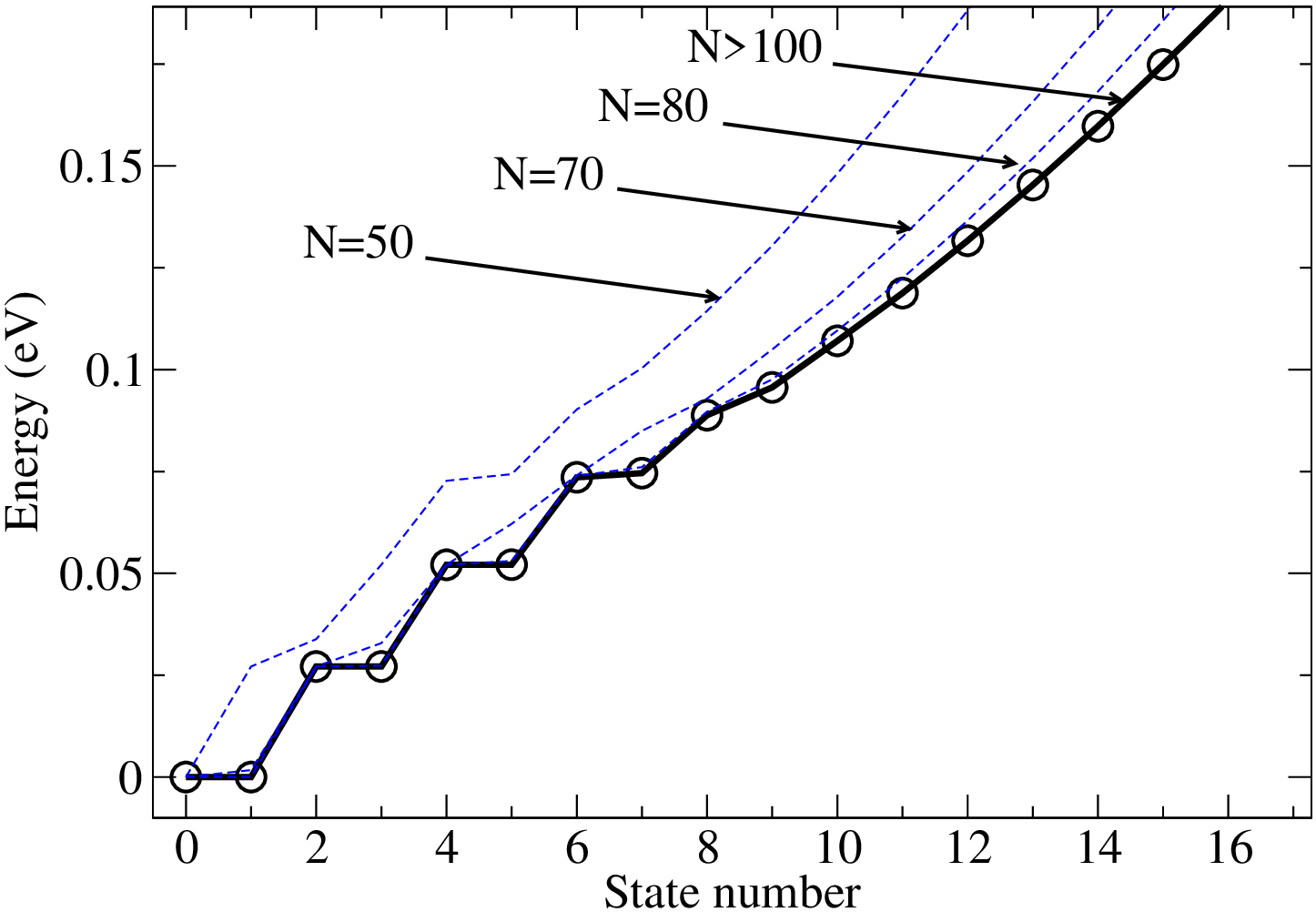}
\caption{\label{fig1} \bf{ (2004) L. Proville}}
\begin{picture}(300,00)(0,0)
\put(-15,350){\makebox(0,0){{\bf (a)}}}
\put(-15,60){\makebox(0,0){{\bf (b)} }}
\end{picture}\end{figure}

{Fig. \ref{fig1}: Energy spectrum of a single He atom embedded
into a double-well potential. The rank $\alpha$ of each eigenvalue
$\gamma(\alpha)$ is given on the X-axis. (a) The semi-classical
calculation (square symbols, dashed line) is compared to the
Hamiltonian diagonalization (circle symbols, solid line) 
onto the truncated Einstein basis (see text below
Eq.\ref{DBa}) with a cutoff $N=100$. The insert shows the
double-well plot versus the displacement (in $\AA$). (b) The latter
method is shown to converge to steady eigenvalues when
the basis cutoff $N$ increases $(N_\infty=2000)$.}\\

%{Fig. \ref{fig2}:  Plot of the bracket
%$<\phi_{\alpha,i}|X_i|\phi_{0,i}>$ for the nonlinear parameters
%$A_4=0.2$ and $A_3=0$ (solid line with triangle symbols) and
%$A_4=0.01$ and $A_3=0.13$ (dashed line with circle symbols)}\\

%{Fig. \ref{fig3}: (color online) For a unitless coupling $C=0.1$,
%comparison between the analytical calculation (dark solid line)
%and our numerics (see in Sec. \ref{Computing}), for the
%Hamiltonian eigenvalues of a 1D harmonic chain, composed of $S=25$
%atoms. The energy-spectrum is arranged in increasing order and it
%is measured with respect to the groundstate. The numerics are
%performed for different upper number of onsite excitations
%$N_{cut}=4$ (blue solid line, circle symbols), $N_{cut}=3$ (red
%dotted line, squares) and $N_{cut}=2$ (green dot-dashed line,
%triangles).}\\

\begin{figure}
\noindent
\includegraphics[width= 16cm]{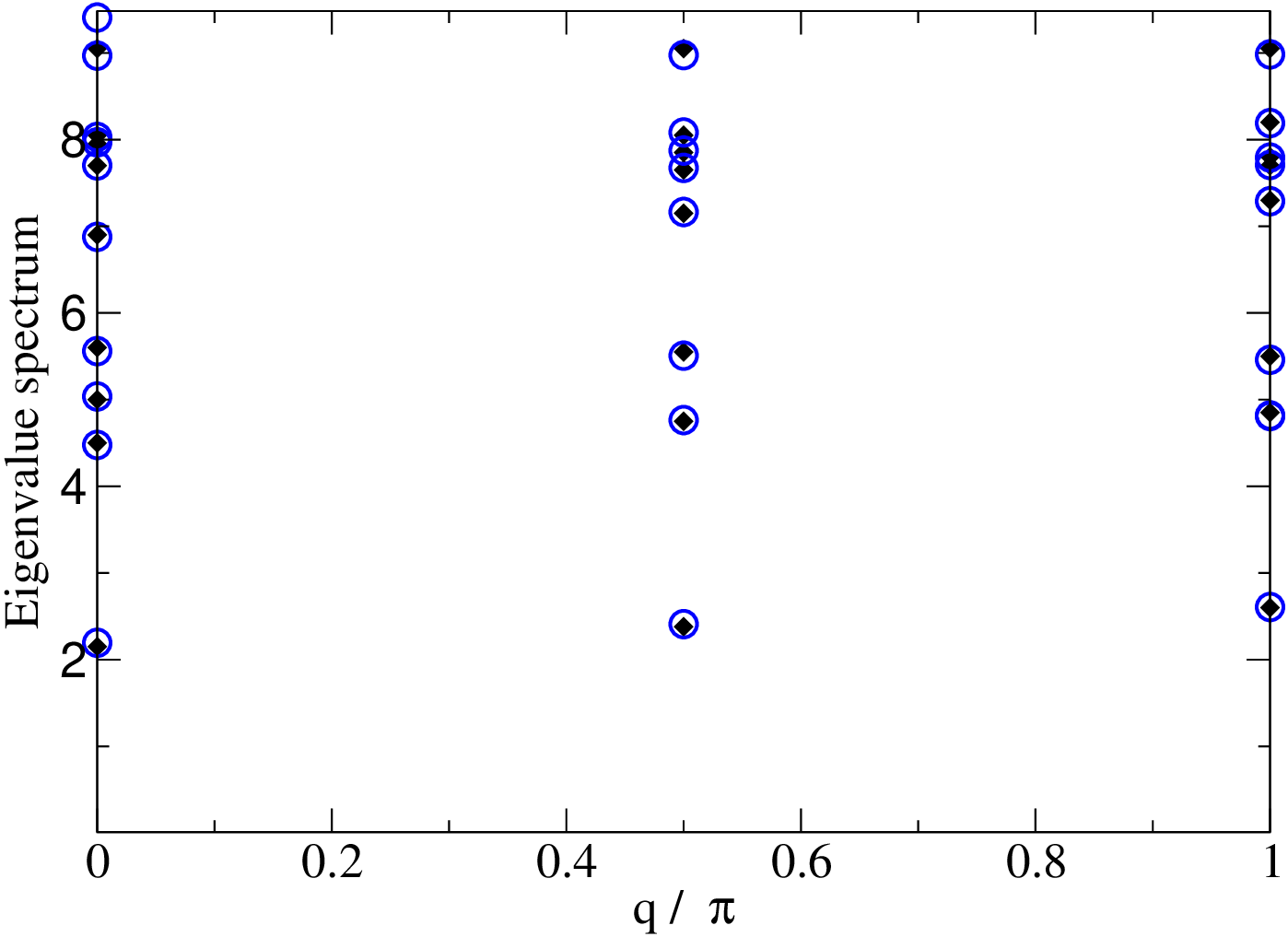}\\
\caption{\label{fig4} \bf{ (2004) L. Proville}}
\end{figure}

\begin{figure}
\noindent
\includegraphics[width= 11.5cm]{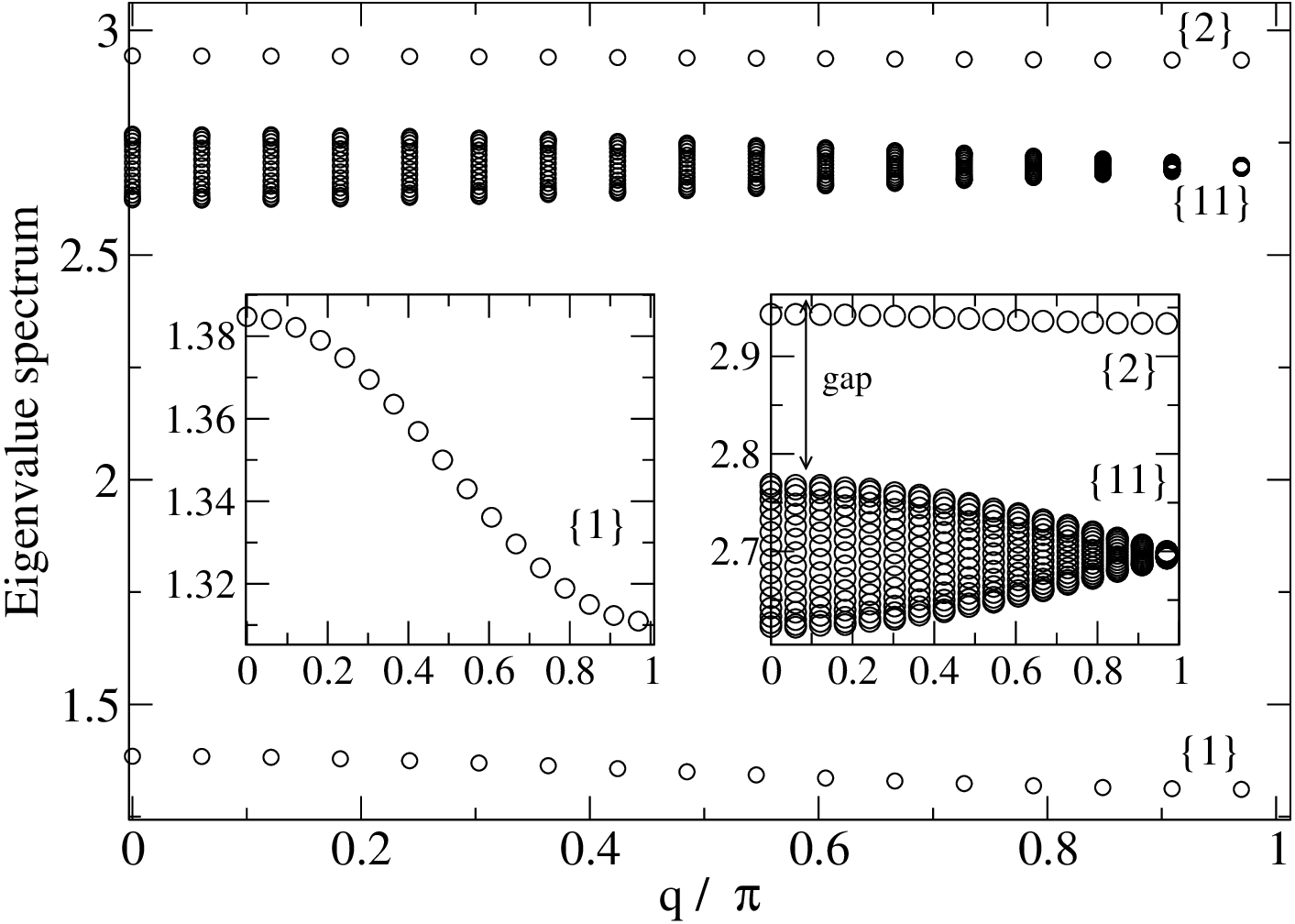}\\
\vspace{1.9cm}
\includegraphics[width= 11.5cm]{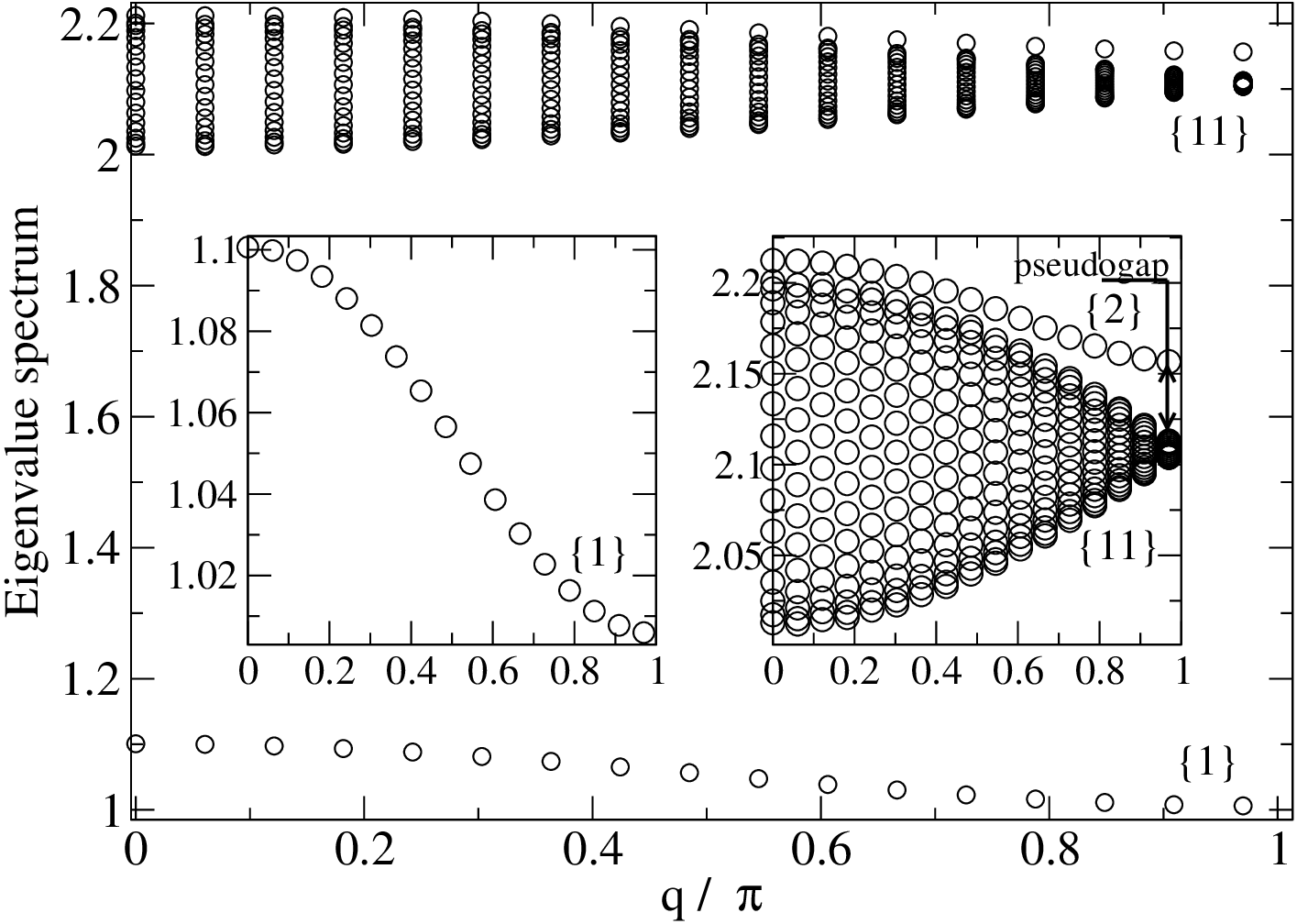}
\caption{\label{fig5} \bf{ (2004) L. Proville}}
\begin{picture}(300,100)(0,0)
\put(-20,440){\makebox(0,0){{\bf (a)}}}
\put(-20,150){\makebox(0,0){{\bf (b)} }}
\end{picture}\end{figure}

\begin{figure}
\noindent
\includegraphics[width= 11.5cm]{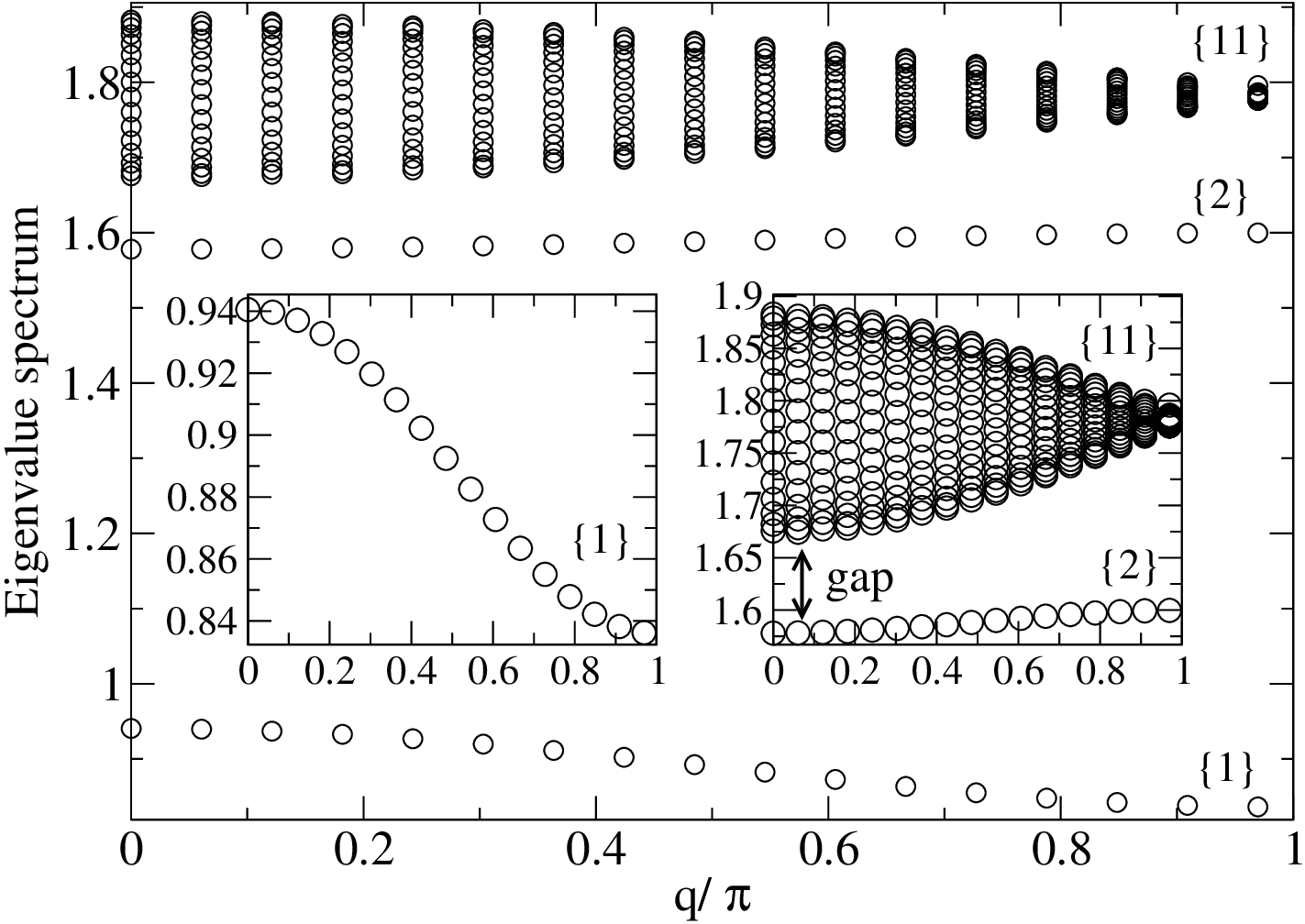}\\
\vspace{1.7cm}
\includegraphics[width= 11.5cm]{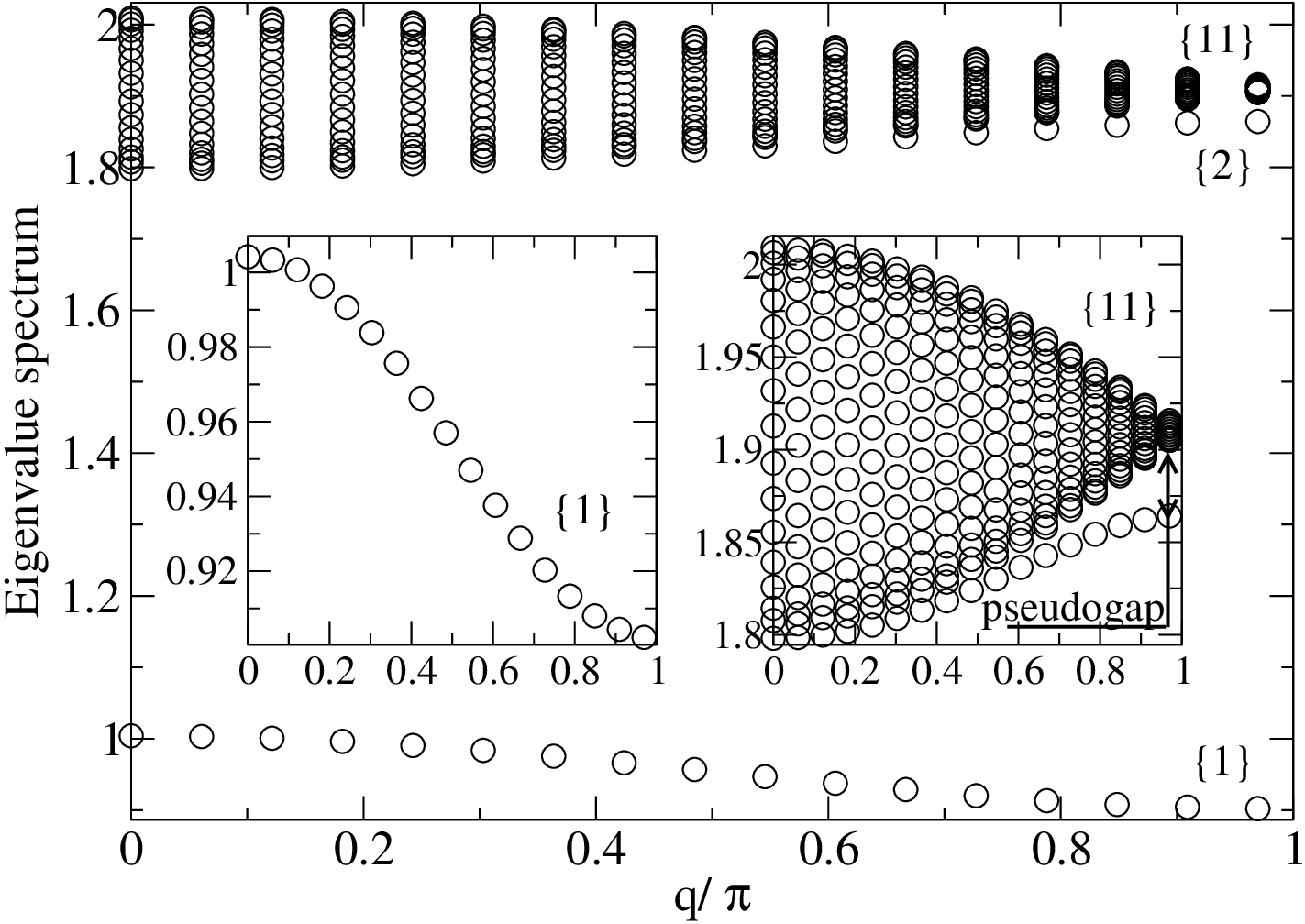}
\caption{\label{fig6} \bf{ (2004) L. Proville}}
\begin{picture}(300,100)(0,0)
\put(-20,430){\makebox(0,0){{\bf (a)}}}
\put(-20,160){\makebox(0,0){{\bf (b)} }}
\end{picture}\end{figure}

{Fig. \ref{fig4}: Eigen-spectrum of a $S=4$ sites chain, with a
$\phi^4$ onsite potential. Model parameters are given in
Ref.~\onlinecite{Param2}. Comparison between our numerics (circle
symbols) and Ref.~\onlinecite{bishop96} (diamond symbols) (see
Sec. \ref{Computing}). The Y-axis unit is $\hbar \Omega$ and its
zero is the groundstate
energy. The X-axis bears the dimensionless first Brillouin zone.}\\

{Fig. \ref{fig5}: Eigen-spectrum of a $S=33$ sites chain for
parameters: (a) $C=0.05$, $A_3=0$ and $A_4=0.2$ and (b) $C=0.05$,
$A_3=0$ and $A_4=0.02$. The inserts show magnifications of the
fundamental branch (left) and the overtone region (right). The
biphonon branch is marked by ($\{2\}$) and the two-phonon band by
($\{11\}$). The same tags label the inserts. The axis units
are same as in Fig.~\ref{fig4}.}\\

{Fig. \ref{fig6}: Same as Fig.~\ref{fig5} but for different
nonlinear parameters: (a) $A_4=0.01$ and $A_3=0.13$  and (b)
$A_4=0.01$ and $A_3=0.105$.}\\

{Fig. \ref{fig7}: Plot of the energy spectrum of a 1D chain,
composed of $S=19$ atoms for: (a) $A_4=0.2, A_3=0.$ and (b)
$A_4=0.01$, $A_3=0.13$, versus the dimensionless coupling $C$. Tags are explained in the text.}\\

{Fig. \ref{fig10}: (color online) On the left hand side, energy
spectrum of a 2D square lattice in the region of the biphonon for
$C=0.025$, $A_3=0$ and $A_4=0.1$ (top) and for same parameters but
$A_4=0.025$ (bottom). The lattice size is $S=13 \times 13$. On the
right hand side, profiles of the left-hand side spectra along the
direction $[11]$. The inserts show the magnifications of
the phonon branch (left) and the biphonon energy region (right).}\\

{Fig. \ref{fig12}: (color online) Plot of the correlation
functions $f(\Phi,n)$ for 3 eigenstates of a $S=23$ chain.
Parameters are $A_4=0.2$, $A_3=0$, $C=0.05$, versus the
dimensionless distance $n$. The eigenstates are (a) phonon states
, (b) the biphonon and (c) the triphonon with for each of them
different wave vectors $q=0$ (circles), $q=4 \pi/S$ (triangles),
$q=10 \pi/S$ (diamonds) , $q=16 \pi/S$ (squares) and $q=22 \pi/S$
(triangles). The inserts in (b) and (c) show a
magnification of the zero y-axis.}\\

{Fig.\ref{fig15}: (color online) Same as Fig.~\ref{fig12} but for
$A_4=0.01$, $A_3=0.13$ and $C=0.05$.}\\

{Fig. \ref{fig13}: (color online) Same as Fig.\ref{fig12} but for
parameters $A_4=0.02$, $A_3=0.$ and $C=0.05$. The eigenstates are
(a) the phonon states and (b) the
biphonon.}\\

{Fig.\ref{fig18}: Plot of the scalar product  $|<\psi_\alpha
(q)|\Phi_\alpha (q)>|^2$ (see Sec.~\ref{Correlations}) for
$C=0.05$ and $A_3=0$ and for $q=0$ (solid lines) and $q=\pi$
(dashed lines), versus the dimensionless parameter $A_4$. The
eigenstates are (a) the phonon states, (b) the biphonon states, and (c) the triphonon states.}\\

\newpage

%%%%%%%%%%%%%%%%%%%%%%%%%%%%%%%%%%%%%%%%%%%%%%%%%%%%%%%%%%%%
%%%%%%%%%%%%%%%%%%%%%%%%%%%%%%%%%%%%%%%%%%%%%%%%%%%%%%%%%%%%
%\newpage
%%%%%%%%%%%%%%%%%%%%%%%%%%%%%%%%%%%%%%%%%%%%%%%%%%%%%%%%%%%%
%\begin{figure}
%\noindent
%\includegraphics[width= 14cm]{fig2PRB2004PROVILLE.eps}
%\caption{\label{fig2} \bf{ (2004) L. Proville}}
%\end{figure}
%%%%%%%%%%%%%%%%%%%%%%%%%%%%%%%%%%%%%%%%%%%%%%%%%%%%%%%%%%%%
\newpage
%%%%%%%%%%%%%%%%%%%%%%%%%%%%%%%%%%%%%%%%%%%%%%%%%%%%%%%%%%%%
%\begin{figure}
%\noindent
%\includegraphics[width= 16cm]{fig3PRB2004PROVILLE.eps}
%\caption{\label{fig3} \bf{ (2004) L. Proville}}
%\end{figure}
%%%%%%%%%%%%%%%%%%%%%%%%%%%%%%%%%%%%%%%%%%%%%%%%%%%%%%%%%%%%
%\newpage
%%%%%%%%%%%%%%%%%%%%%%%%%%%%%%%%%%%%%%%%%%%%%%%%%%%%%%%%%%%%
%%%%%%%%%%%%%%%%%%%%%%%%%%%%%%%%%%%%%%%%%%%%%%%%%%%%%%%%%%%%
\newpage
%%%%%%%%%%%%%%%%%%%%%%%%%%%%%%%%%%%%%%%%%%%%%%%%%%%%%%%%%%%%
%%%%%%%%%%%%%%%%%%%%%%%%%%%%%%%%%%%%%%%%%%%%%%%%%%%%%%%%%%%%
\newpage
%%%%%%%%%%%%%%%%%%%%%%%%%%%%%%%%%%%%%%%%%%%%%%%%%%%%%%%%%%%%
%%%%%%%%%%%%%%%%%%%%%%%%%%%%%%%%%%%%%%%%%%%%%%%%%%%%%%%%%%%%
\newpage
%%%%%%%%%%%%%%%%%%%%%%%%%%%%%%%%%%%%%%%%%%%%%%%%%%%%%%%%%%%%
\begin{figure}
\noindent
\vspace{1.7cm}
\caption{\label{fig7} \bf{(2004) L. Proville: See additional figures}}
%\begin{picture}(300,100)(0,0)
%\put(15,380){\makebox(0,0){{\bf (a)}}}
%\put(15,160){\makebox(0,0){{\bf (b)} }}
%\end{picture}
\end{figure}

%%%%%%%%%%%%%%%%%%%%%%%%%%%%%%%%%%%%%%%%%%%%%%%%%%%%%%%%%%%%
\newpage
%%%%%%%%%%%%%%%%%%%%%%%%%%%%%%%%%%%%%%%%%%%%%%%%%%%%%%%%%%%%
\begin{figure}
\noindent
\includegraphics[width= 8.5cm,height=8cm]{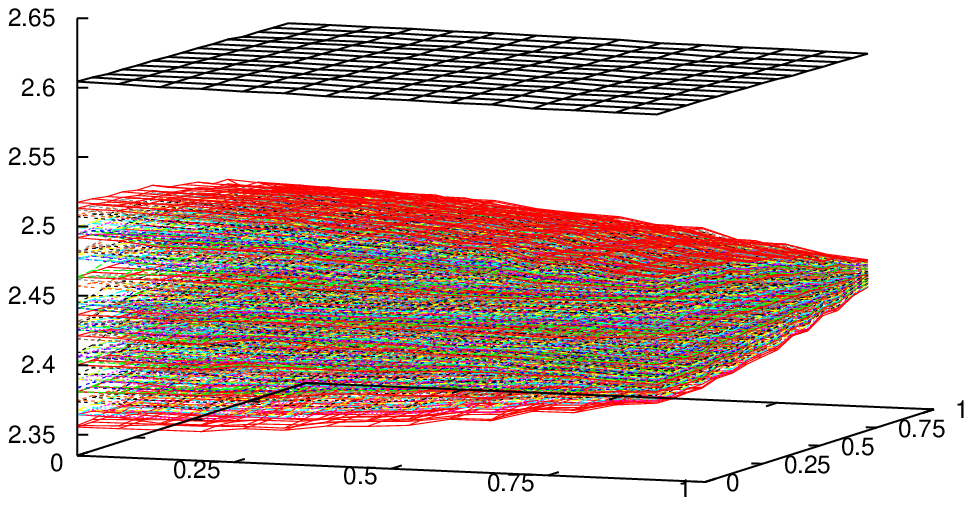}
\includegraphics[width= 7.5cm]{fig9bPRB2004PROVILLE.eps}\\
\vspace{1cm}
\includegraphics[width= 8.5cm,height=8cm]{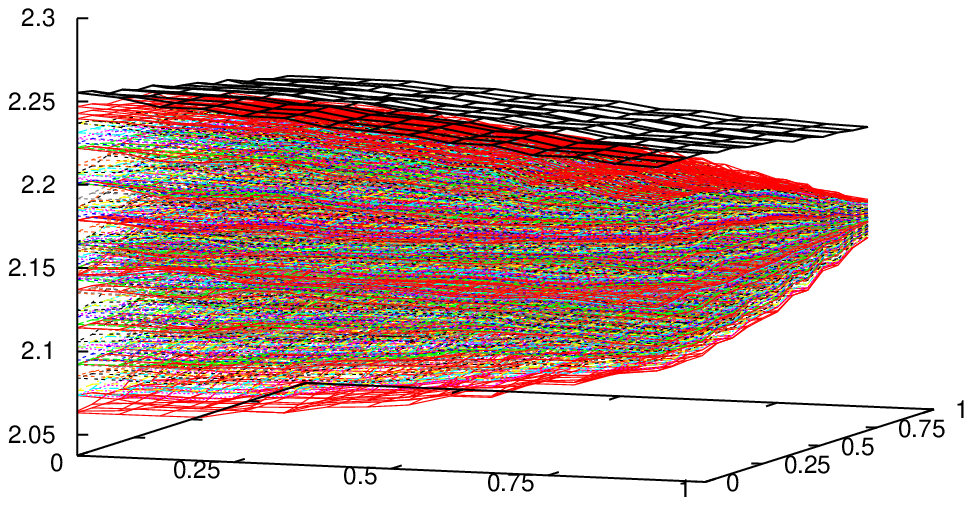}
\includegraphics[width= 7.5cm]{fig9dPRB2004PROVILLE.eps}
\caption{\label{fig10} \bf{ (2004) L. Proville}}
\begin{picture}(300,260)(0,0)
\put(120,605){\makebox(0,0){ $q_y / \pi$}}
\put(30,590){\makebox(0,0){ $q_x / \pi$}}
\put(120,347){\makebox(0,0){ $q_y / \pi$}}
\put(30,336){\makebox(0,0){ $q_x / \pi$}}
\put(-60,615){\begin{rotate}{90}
{\small 2D lattice eigenvalue}
\end{rotate}}
\put(-60,360){\begin{rotate}{90}
{\small 2D lattice eigenvalue}
\end{rotate}}
\end{picture}
\end{figure}
%%%%%%%%%%%%%%%%%%%%%%%%%%%%%%%%%%%%%%%%%%%%%%%%%%%%%%%%%%%%

\newpage

%%%%%%%%%%%%%%%%%%%%%%%%%%%%%%%%%%%%%%%%%%%%%%%%%%%%%%%%%%%%
\begin{figure}
\noindent
\includegraphics[height=5cm]{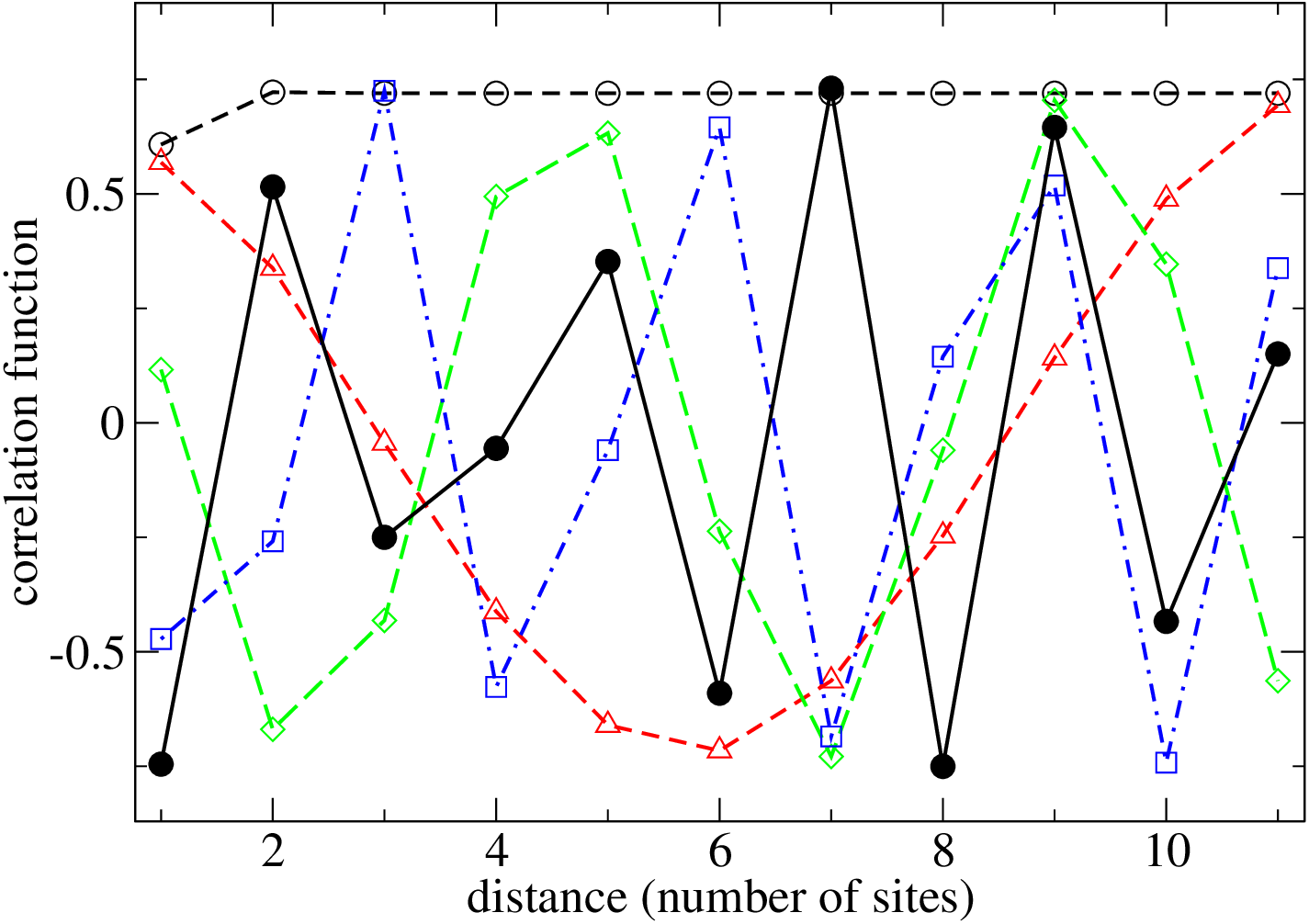}\\
\vspace{1cm}
\includegraphics[height=5cm]{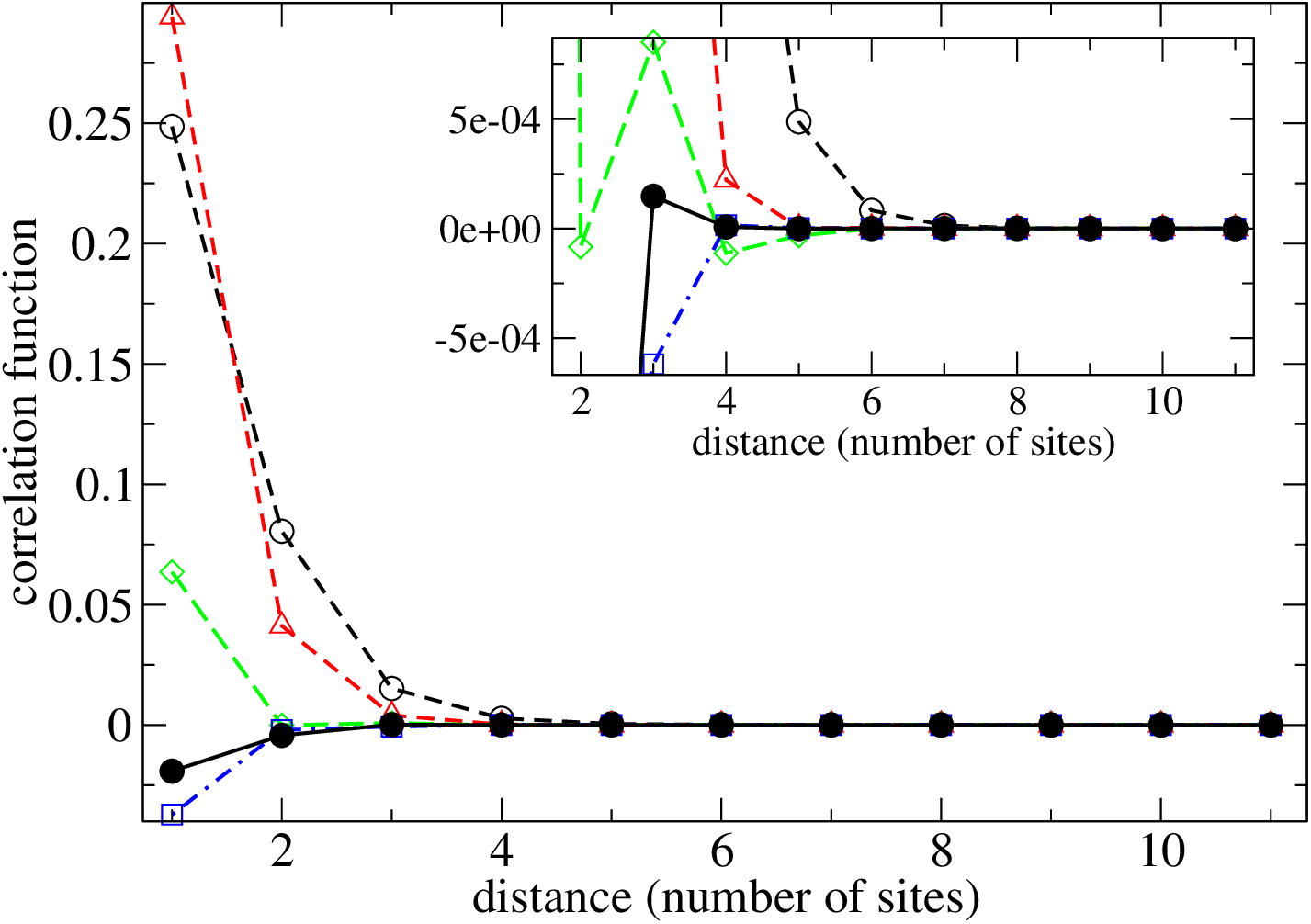}\\
\vspace{1cm}
\includegraphics[height=5cm]{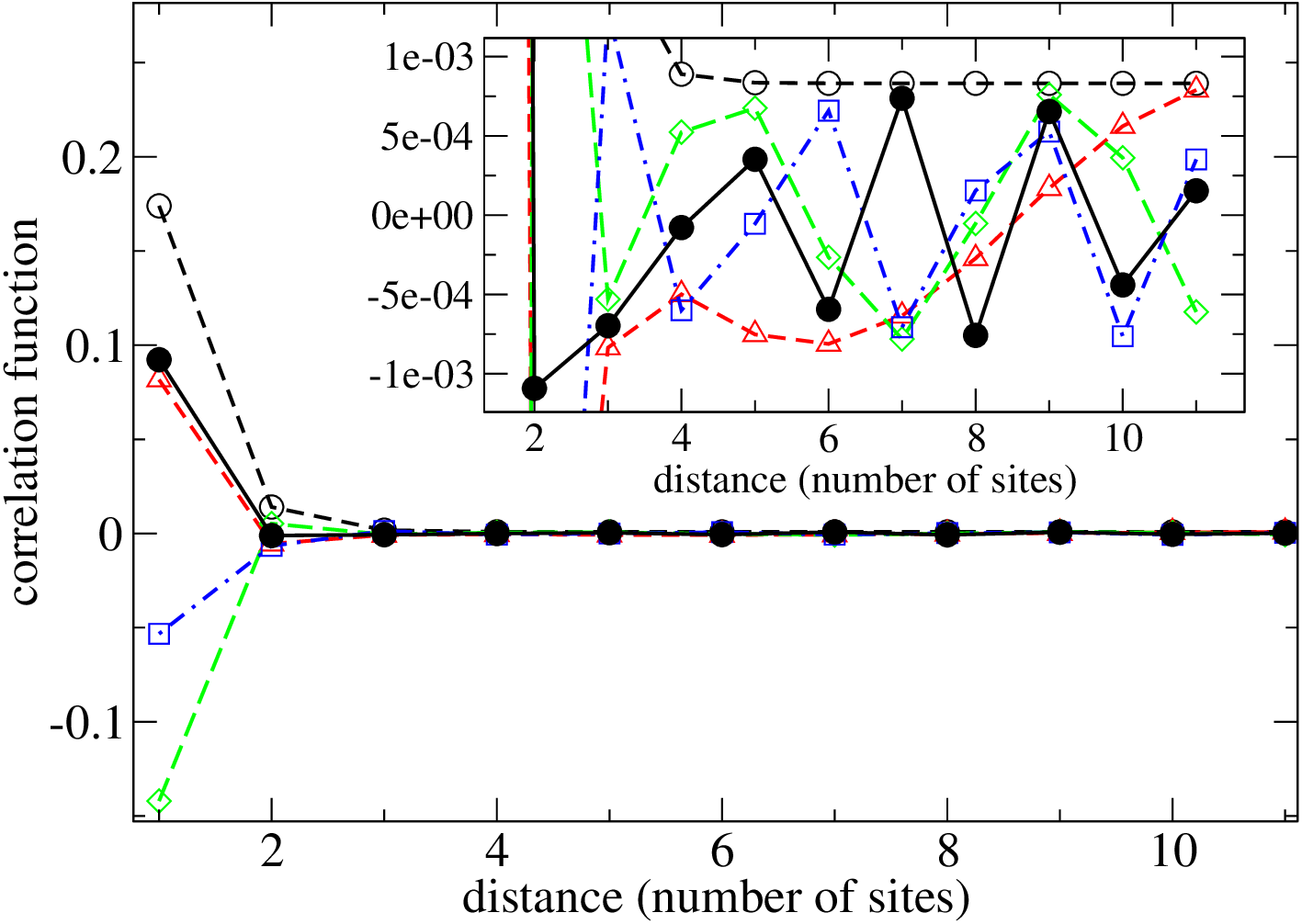}
\caption{\label{fig12} \bf{ (2004) L. Proville}}
\begin{picture}(300,400)(0,0)
\put(15,790){\makebox(0,0){{\bf (a)}}}
\put(15,620){\makebox(0,0){{\bf (b)} }}
\put(15,450){\makebox(0,0){ {\bf (c)} }}
\end{picture}
\end{figure}
%%%%%%%%%%%%%%%%%%%%%%%%%%%%%%%%%%%%%%%%%%%%%%%%%%%%%%%%%%%%

\newpage

%%%%%%%%%%%%%%%%%%%%%%%%%%%%%%%%%%%%%%%%%%%%%%%%%%%%%%%%%%%%
\begin{figure}
\noindent
\includegraphics[height=5cm]{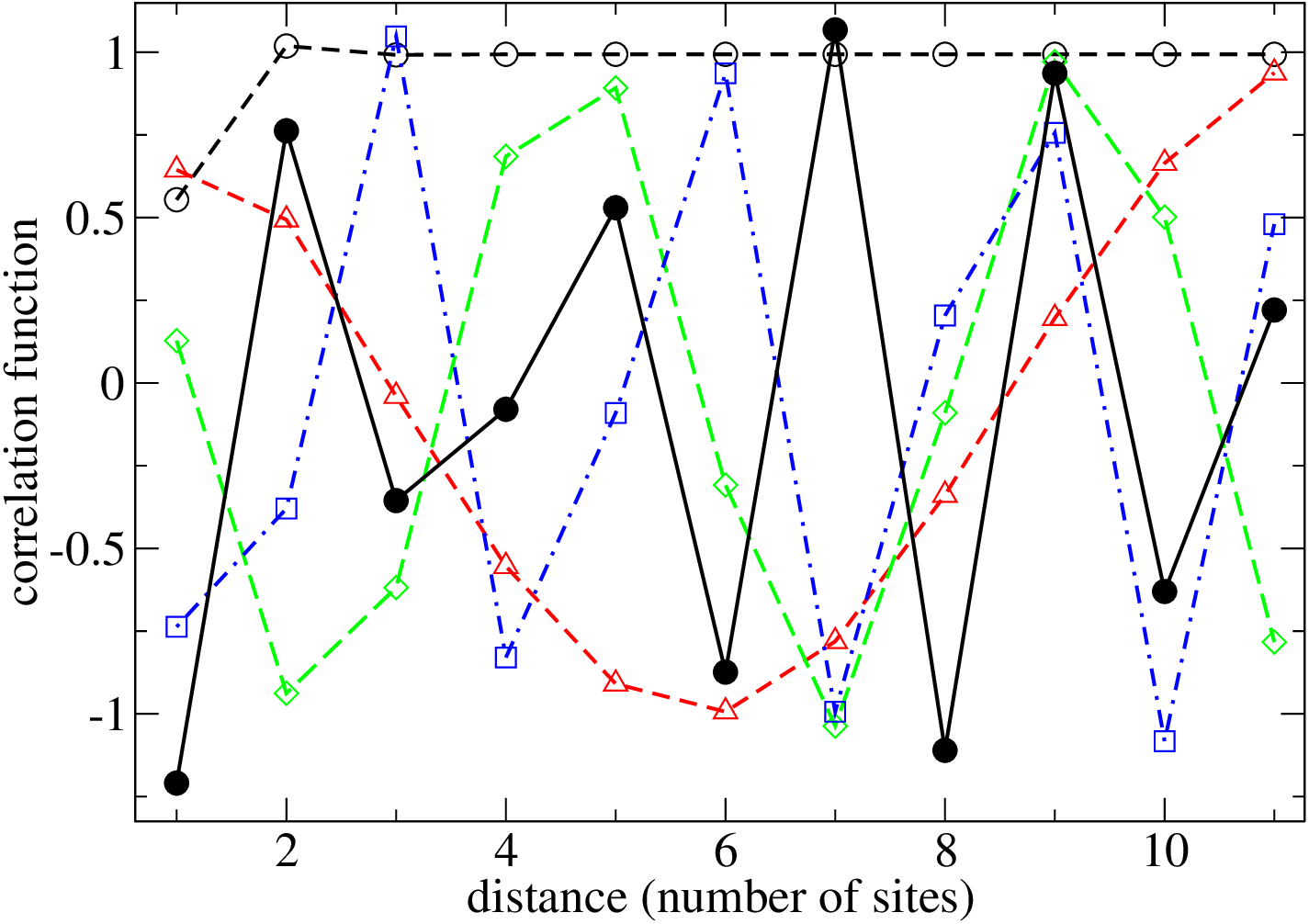}\\
\vspace{1cm}
\includegraphics[height=5.cm]{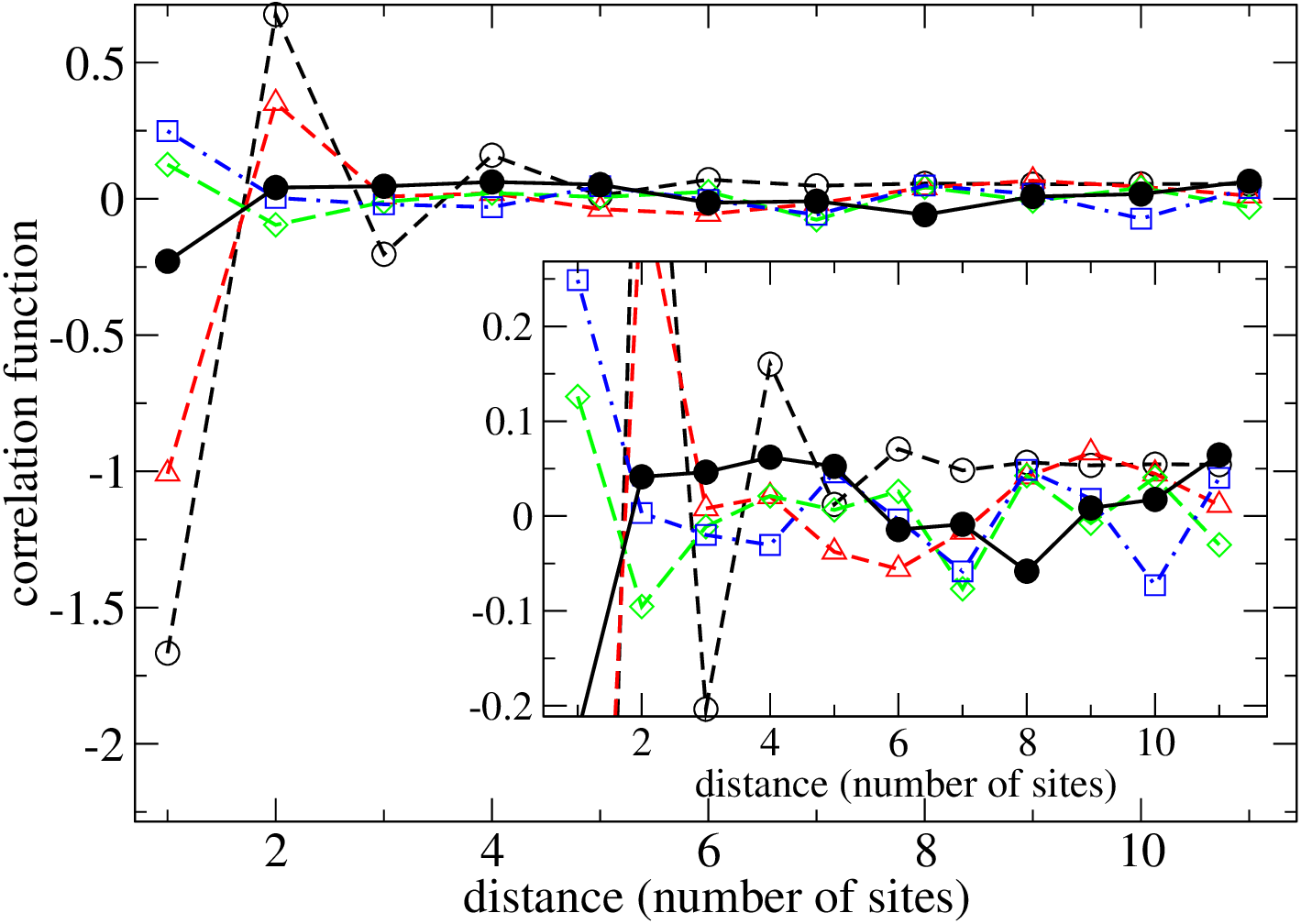}\\
\vspace{1cm}
\includegraphics[height=5.cm]{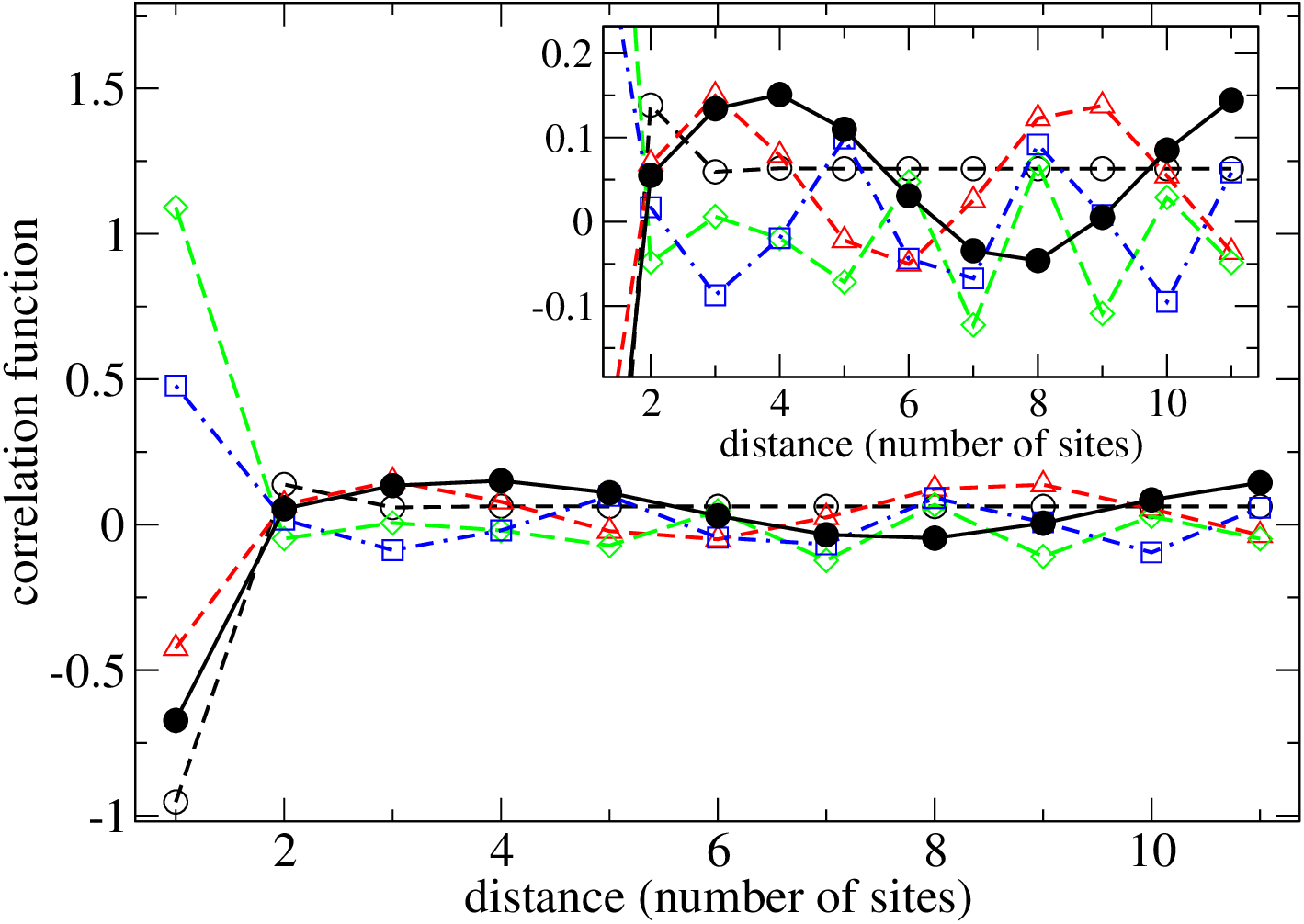}
\caption{\label{fig15} \bf{ (2004) L. Proville}}
\begin{picture}(300,400)(0,0)
\put(15,790){\makebox(0,0){{\bf (a)}}}
\put(15,620){\makebox(0,0){{\bf (b)} }}
\put(15,450){\makebox(0,0){{\bf (c)} }}
\end{picture}
\end{figure}
%%%%%%%%%%%%%%%%%%%%%%%%%%%%%%%%%%%%%%%%%%%%%%%%%%%%%%%%%%%%
\newpage

%%%%%%%%%%%%%%%%%%%%%%%%%%%%%%%%%%%%%%%%%%%%%%%%%%%%%%%%%%%%
\begin{figure}
\noindent
\includegraphics[height=5.cm]{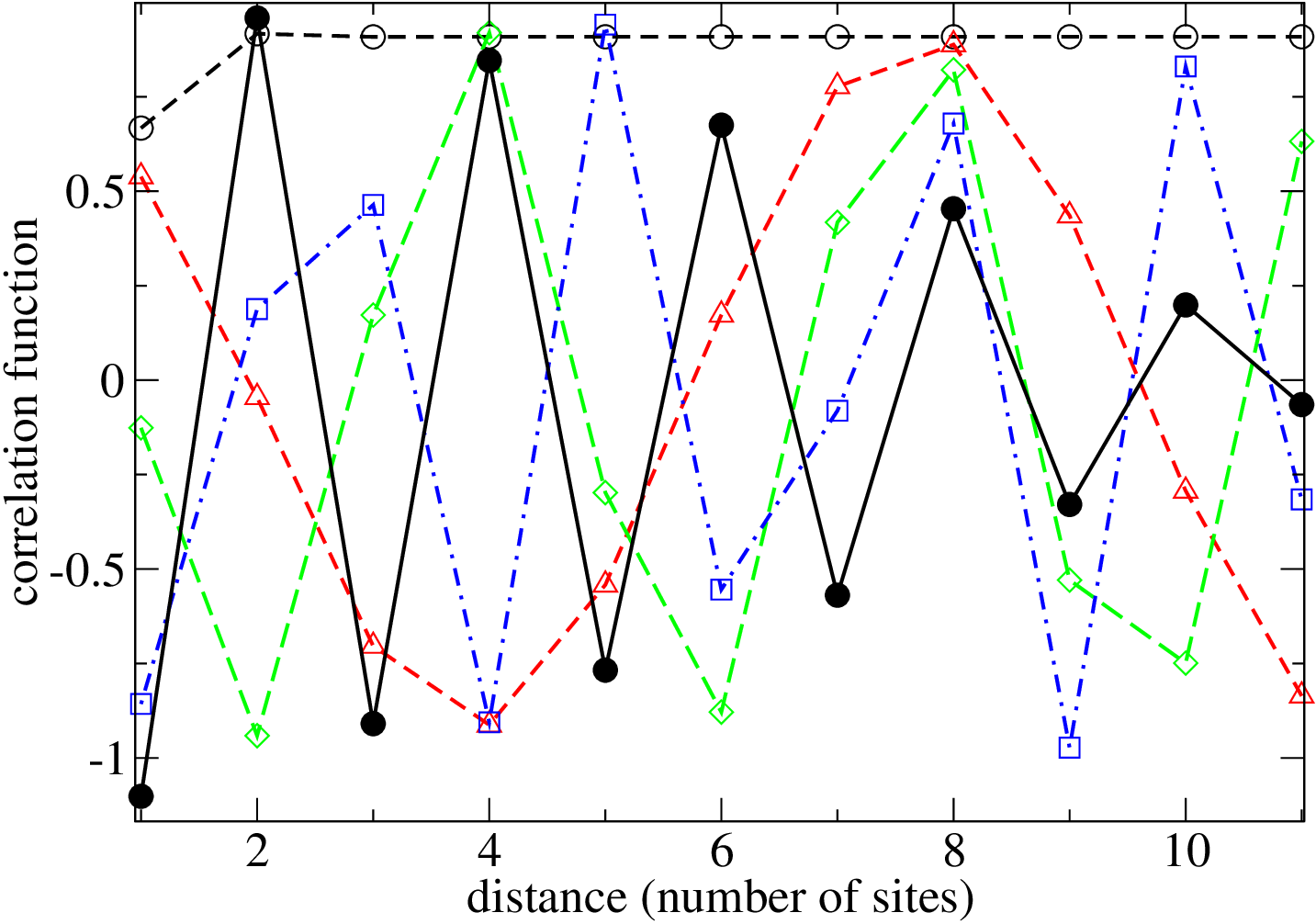}\\
\vspace{1cm}
\includegraphics[height=5.cm]{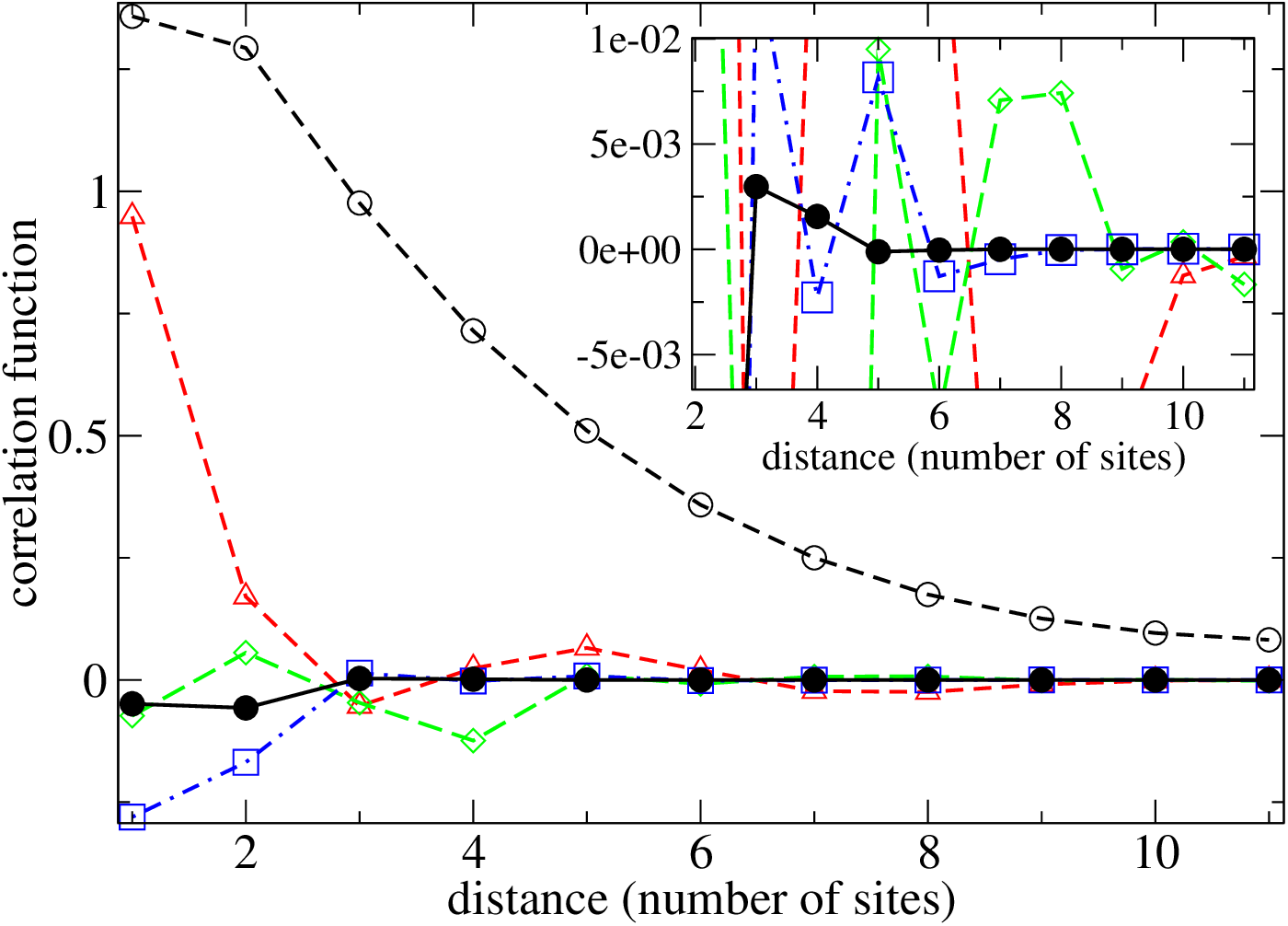}
\caption{\label{fig13} \bf{ (2004) L. Proville}}
\begin{picture}(300,400)(0,0)
\put(15,620){\makebox(0,0){{\bf (a)}}}
\put(15,450){\makebox(0,0){{\bf (b)} }}
\end{picture}
\end{figure}
%%%%%%%%%%%%%%%%%%%%%%%%%%%%%%%%%%%%%%%%%%%%%%%%%%%%%%%%%%%%
\newpage

%%%%%%%%%%%%%%%%%%%%%%%%%%%%%%%%%%%%%%%%%%%%%%%%%%%%%%%%%%%%
%\begin{figure}
%\noindent
%\includegraphics[height=5.cm]{Fig14aPRB2004PROVILLE.eps}\\
%\vspace{1cm}
%\includegraphics[height=5.cm]{Fig14bPRB2004PROVILLE.eps}
%\caption{\label{fig14} \bf{ (2004) L. Proville}}
%\begin{picture}(300,400)(0,0)
%\put(15,620){\makebox(0,0){{\bf (a)}}}
%\put(15,450){\makebox(0,0){{\bf (b)} }}
%\end{picture}
%\end{figure}
%%%%%%%%%%%%%%%%%%%%%%%%%%%%%%%%%%%%%%%%%%%%%%%%%%%%%%%%%%%%

\newpage

%%%%%%%%%%%%%%%%%%%%%%%%%%%%%%%%%%%%%%%%%%%%%%%%%%%%%%%%%%%%
\begin{figure}
\noindent
\includegraphics[height=6.cm]{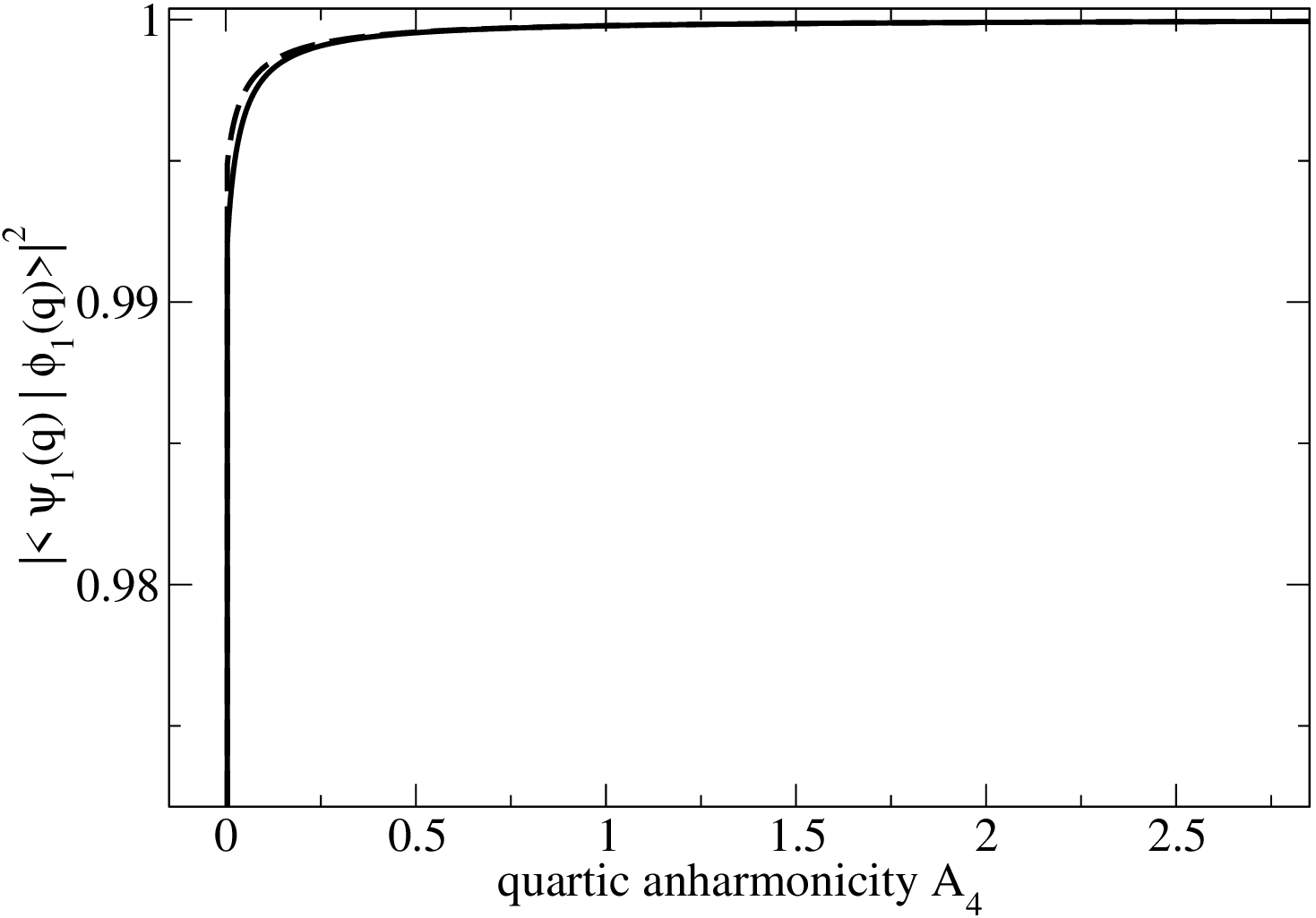}\\
\vspace{1.5cm}
\includegraphics[height=6.cm]{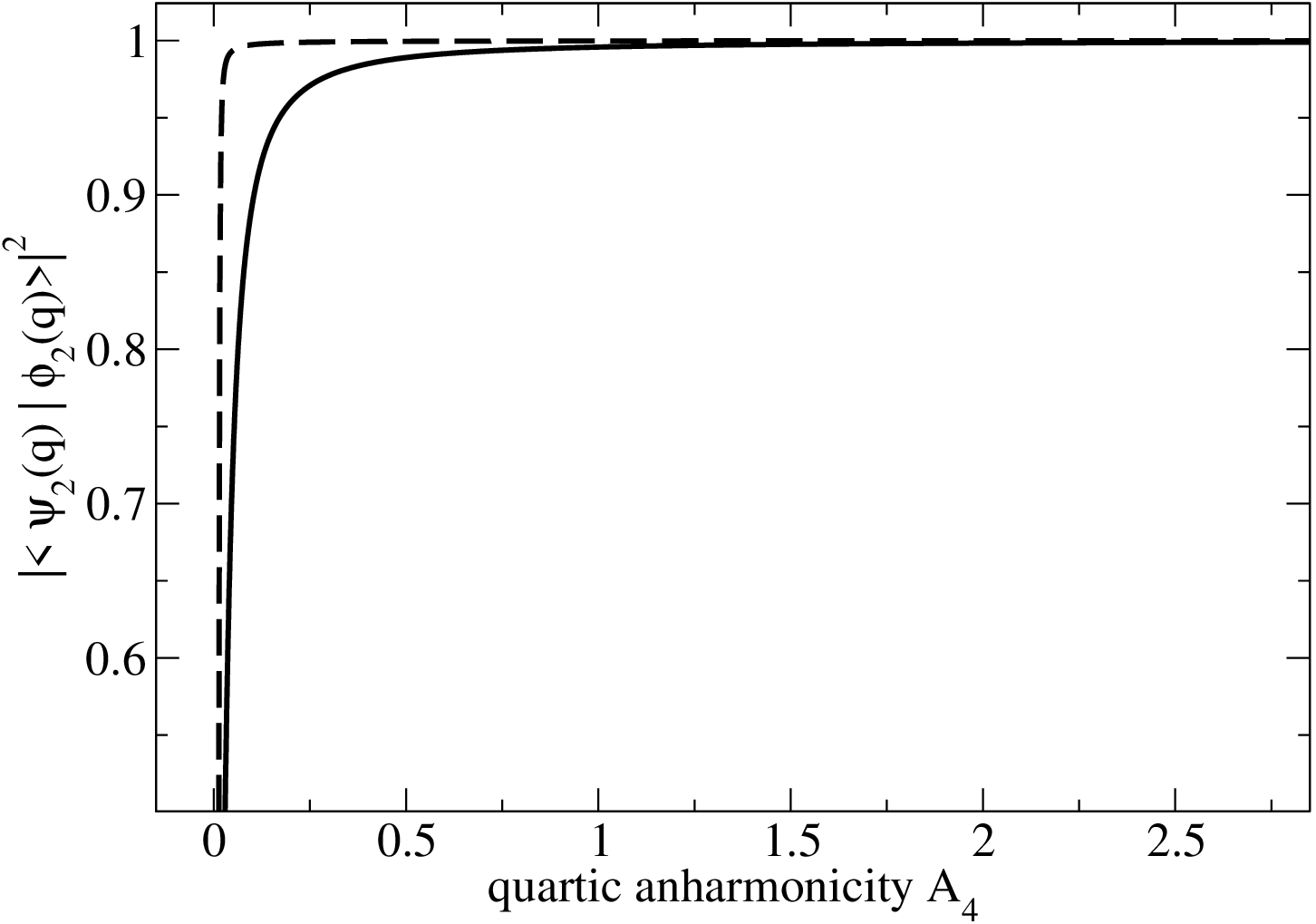}\\
\vspace{1.5cm}
\includegraphics[height=6.cm]{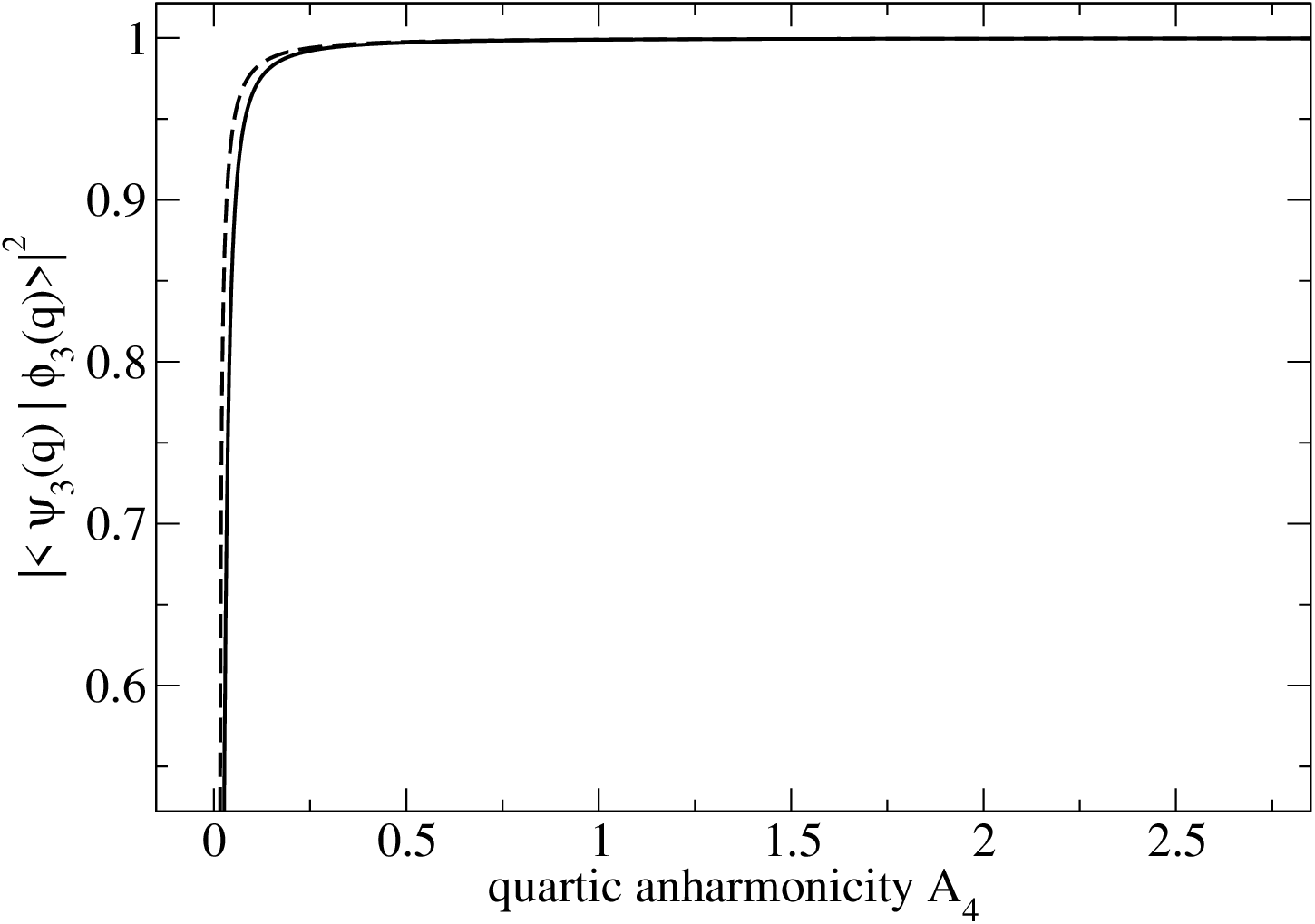}
\caption{\label{fig18} \bf{ (2004) L. Proville}}
\begin{picture}(300,400)(0,0)
\put(15,885){\makebox(0,0){{\bf (a)}}}
\put(15,680){\makebox(0,0){{\bf (b)} }}
\put(15,470){\makebox(0,0){{\bf (c)} }}
\end{picture}\end{figure}
%%%%%%%%%%%%%%%%%%%%%%%%%%%%%%%%%%%%%%%%%%%%%%%%%%%%%%%%%%%%

\newpage

%%%%%%%%%%%%%%%%%%%%%%%%%%%%%%%%%%%%%%%%%%%%%%%%%%%%%%%%%%%%
%\begin{figure}
%\noindent
%\includegraphics[height=8.cm]{fig11aPRB2004PROVILLE.eps}
%\caption{\label{fig11} \bf{ (2004) L. Proville}}
%\begin{picture}(300,400)(0,0)
%\put(15,620){\makebox(0,0){{\bf (a)}}}
%\put(15,450){\makebox(0,0){{\bf (b)} }}
%\end{picture}
%\end{figure}
%{Fig.\ref{fig11}: Plot of the band width ratio $\Delta_2/\Delta_1$, i.e.,
%the biphonon branch width divided by the phonon branch width
%versus the two biphonon gaps: the gap above the
%biphonon branch (dashed line) and for the gap below the
%biphonon branch (solid line).
%The parameter $A_4>0$ is varying
%while $A_3=0$ and $C=0.05$.}\\

%%%%%%%%%%%%%%%%%%%%%%%%%%%%%%%%%%%%%%%%%%%%%%%%%%%%%%%%%%%%

\begin{table}
\begin{center}
\begin{tabular} {|c||cc|cc|cc|cc|}
\hline \hline $2 <\phi_{\alpha_i}|X_i|\phi_{0_i}>^2  $ &
\multicolumn{2}{|c|}{$A_4=0.2 \ A_3=0$}&\multicolumn{2}{|c|}{$A_4=0.01 \ A_3=0.13$}&\multicolumn{2}{|c|}{$A_4=0.02\ A_3=0$}&\multicolumn{2}{|c|}{$A_4=0.01\ A_3=0.105$}\\
&Fig.\ref{fig5}(a) \& &Fig.\ref{fig12} & Fig.\ref{fig6}(a) \& &Fig.\ref{fig15}& Fig.\ref{fig5}(b) \& &
Fig.\ref{fig13}& Fig.\ref{fig6}(b) &\\
\hline
$\alpha=1$   &  $0.74$ &  &  $1.06$ &    & $0.95$ &  & $1.03$ & \\
$\alpha=2$   &   $0.$ &   &  $2.97 10^{-2}$ &   &  $0.$ &  &    $1.16 10^{-2}$ & \\
$\alpha=3$   &   $7.39 10^{-4}$ &    &   $2.36 10^{-3}$ &   & $7.9 10^{-5}$ &  &  $3.3 10^{-4}$  & \\
$\alpha=4$   &   $0.$ &    &   $3.91 10^{-4}$  &  & $0.$ &   &    $1.09 10^{-5}$  & \\
$\alpha=5$   &   $8.45 10^{-7}$ &    &   $5.45 10^{-5}$  &  & $1.45 10^{-8}$ &   &    $4.39 10^{-7}$  & \\
$\alpha=6$   &   $0.$ &    &   $6.38 10^{-6}$  &  & $0.$ &   &    $1.91 10^{-8}$  & \\
\hline \hline
\end{tabular}
\caption{ \label{tab} \bf{ (2004) L. Proville}}
\end{center}
\end{table}

% 1.06
% 0.0272
% 1.8


\begin{thebibliography}{99}
\bibitem{Param1} when $E_d=0.1$ eV and $b= 2$ \AA,
the parameters in Eq. (\ref{DBa}) are $A_3=-0.09496$,  $A_4=0.00451$ and
$\hbar \Omega=0.02886 eV$.
\end{thebibliography}
\end{document}